\documentclass[12pt]{article}
\usepackage{graphicx}

\catcode`\@=11
\def\marginnote#1{}

\def\draftlabel#1{{\@bsphack\if@filesw {\let\thepage\relax
   \xdef\@gtempa{\write\@auxout{\string
      \newlabel{#1}{{\@currentlabel}{\thepage}}}}}\@gtempa
   \if@nobreak \ifvmode\nobreak\fi\fi\fi\@esphack}
        \gdef\@eqnlabel{#1}}
\def\@eqnlabel{}
\def\@vacuum{}
\def\draftmarginnote#1{\marginpar{\raggedright\scriptsize\tt#1}}
\def\draft{\oddsidemargin -.5truein
        \def\@oddfoot{\sl preliminary draft \hfil
        \rm\thepage\hfil\sl\today}
        \let\@evenfoot\@oddfoot \overfullrule 3pt
        \let\label=\draftlabel
        \let\marginnote=\draftmarginnote
   \def\@eqnnum{(\theequation)\rlap{\kern\marginparsep\tt\@eqnlabel}%
\global\let\@eqnlabel\@vacuum}  }

\def\preprint{\twocolumn\sloppy\flushbottom\parindent 1em
        \leftmargini 2em\leftmarginv .5em\leftmarginvi .5em
        \oddsidemargin -.5in    \evensidemargin -.5in
        \columnsep 15mm \footheight 0pt
        \textwidth 250mmin      \topmargin  -.4in
        \headheight 12pt \topskip .4in
        \textheight 175mm
        \footskip 0pt
        \def\@oddhead{\thepage\hfil\addtocounter{page}{1}\thepage}
        \let\@evenhead\@oddhead \def\@oddfoot{} \def\@evenfoot{} }

\def\titlepage{\@restonecolfalse\if@twocolumn\@restonecoltrue\onecolumn
     \else \newpage \fi \thispagestyle{empty}\c@page\z@
        \def\thefootnote{\fnsymbol{footnote}} }

\def\endtitlepage{\if@restonecol\twocolumn \else  \fi
        \def\thefootnote{\arabic{footnote}} \setcounter{footnote}{0}}

\catcode`@=12 \relax
\def\be{\begin{equation}}
\def\ee{\end{equation}}
\def\ba{\begin{eqnarray}}
\def\ea{\end{eqnarray}}

\def\Im{\mathop{\rm Im}}
\def\Re{\mathop{\rm Re}}
\def\d{\partial}

\def\cM{{\cal M}}
\def\intm{\int_{\cM} {\rm d}^4 x\,}
\def\intdm{\int_{\partial\cM} {\rm d}^3 x\,}
\def\intdmh{\int_{\partial\cM}\hskip-1.5mm {\rm d}^3 x\,}
\def\intt{\int {\rm d}^2\theta\,}
\def\inttb{\int {\rm d}^2\bar\theta\,}
\def\intttb{\int {\rm d}^2\theta {\rm d}^2\bar\theta\,}

\def\a{\alpha}
\def\b{\beta}
\def\g{\gamma}
\def\G{\Gamma}
\def\dd{\delta}
\def\e{\epsilon}
\def\m{\mu}
\def\n{\nu}

\def\l{\lambda}

\def\s{\sigma}
\def\f{\phi}
\def\p{\psi}
\def\c{\chi}
\def\t{\theta}
\def\tb{\bar\theta}
\def\D{\Delta}
\def\z{\zeta}
\def\w{\omega}
\def\vf{\varphi}

\def\wh{\widehat}
\def\wt{\widetilde}
\def\eb{\bar\epsilon}
\def\pb{\bar\psi}
\def\cb{\bar\chi}
\def\zb{\bar z}
\def\ub{\bar u}
\def\hz{\widehat z}

\def\hp{\widehat \p}

\def\hf{\widehat f}

\def\sb{\bar \sigma}
\def\fb{\bar f}

\def\xb{\overline\xi}

\def\ad{{\dot\a}}
\def\bd{{\dot\b}}
\def\gd{{\dot\g}}

\def\ov{\overline}

\def\rd{\sqrt{2}}
\def\half{{1\over 2}\,}
\def\mh{{\hat\m}}
\def\nh{{\hat\n}}

\def\cG{{\cal G}}
\def\cK{{\cal K}}

\renewcommand{\theequation}{\thesection.\arabic{equation}}

\begin{document}

\topmargin-2.0cm
%
%
%
%
\begin{titlepage}
\begin{flushright}
LPTENS-11/10\\
March 2011\\
\end{flushright}
\vskip 1.5cm

\begin{center}{\Large\bf 
Supersymmetric Boundaries and Junctions\\ in Four Dimensions} 
\vskip 1.0cm 
{\bf Adel Bilal}

\vskip.3cm 
Laboratoire de Physique Th\'eorique,
\'Ecole Normale Sup\'erieure - CNRS\footnote{
Unit\'e mixte du CNRS et de l'Ecole Normale Sup\'erieure associ\'ee \`a l'Universit\'e Paris 6 Pierre et Marie Curie
}\\
\vskip-2.mm
24 rue Lhomond, 75231 Paris Cedex 05, France

\end{center}
\vskip 1.0cm

\setlength{\baselineskip}{.525cm}
\begin{center}
{\bf Abstract}
\end{center}
\begin{quote}
We make a comprehensive study of (rigid) ${\cal N}=1$ supersymmetric sigma-models with general K\"ahler potentials $K$ and superpotentials $w$ on four-dimensional space-times with boundaries. We  determine the minimal (non-supersymmetric) boundary terms one must add to the standard bulk action to make it off-shell invariant under half the supersymmetries without imposing any boundary conditions. Susy boundary conditions do arise from the variational principle when studying the dynamics. Upon including an additional boundary action that depends on an arbitrary real boundary potential $B$ one can generate very general susy boundary conditions. We show that for any set of susy boundary conditions that define a Lagrangian submanifold of the K\"ahler manifold, an appropriate boundary potential $B$ can be found. Thus the non-linear sigma-model on a manifold with boundary is characterised by the tripel $(K,B,w)$. We also discuss  the susy coupling to new boundary superfields and generalize our results to supersymmetric junctions between completely different susy sigma-models,  living on adjacent domains and interacting through a ``permeable'' wall. We obtain the supersymmetric matching conditions that allow us to couple models with different K\"ahler potentials and superpotentials on each side of the wall.  
\end{quote}


\end{titlepage}

\setcounter{footnote}{0} \setcounter{page}{0}
\setlength{\baselineskip}{.525cm}
\newpage
{\small{
\baselineskip=0.5cm\tableofcontents}}
\setcounter{page}{0}
\newpage
%
%
%
\section{Introduction and summary\label{Intro}}
\setcounter{equation}{0}


The study of supersymmetry on spaces with  boundaries has attracted considerable attention in recent years  for a variety of  reasons. An early general study in various dimensions appeared in \cite{early}.  Much of the work has focused on two-dimensional sigma-models \cite{sigma2d,Koerb}, and was  motivated  by the relation of  supersymmetric  boundary conditions with D-branes  \cite{Hori}  and  by the search for  integrability \cite{Warner}.  Other works,  in higher-dimensional space-times, were initiated by the Horava-Witten construction of the heterotic string \cite{HW}.   Maintaining half of the $d=11$  supersymmetry in this construction is,  in particular, crucial in order to  correctly identify  the anomaly-cancelling  boundary contributions to the Bianchi identity, see e.g. \cite{BDSBM}.  The reduction of the Horava-Witten theory on a Calabi-Yau manifold, as well as more general  brane-world constructions in $d=5$ dimensions, 
have been extensively discussed  (for a partial  list of references see \cite{D5}). Finally, the case  $d=4$ has been analyzed mostly  in the context of defect conformal field theories \cite{Erd}, which are  dual to ${\rm AdS}_n$ branes living in a ${\rm AdS}_{n+1}$ bulk  \cite{KR,BDDO}.  A recent discussion of four-dimensional supergravity with boundaries is given in \cite{PvN}, and earlier results on the quantization of Euclidean gravity and supergravity with boundaries are collected in \cite{Esposito}. An interesting study of boundary terms and boundary conditions in rigid 3 and 4 dimensional supersymmetry \cite{PvN2} has also appeared recently.  Boundary conditions in $N=4$ super Yang-Mills theory have been studied in \cite{GaiottoW}. Nevertheless, a systematic analysis of the effect of boundaries in the general $N=1$  sigma-model in $3+1$ dimensions with arbitrary K\"ahler potentials and superpotentials is, to the best of our knowledge,  
still lacking.  Providing such an analysis is the aim of the present work. One of our goals is to determine all consistent, supersymmetric, generally non-linear boundary conditions. As an application, this will allow us to study the possible couplings of different sigma-models on adjacent domains through permeable walls.

A possible ``physical" realization of the latter scenario could be the effective low-energy physics resulting from  a domain wall of a heavy superfield ``that has been integrated out". As an example of what we have in mind, consider a simple Wess-Zumino model consisting of a heavy chiral superfield $\Phi_{\rm H}$ and a light one $\Phi$ with a superpotential 
$W(\Phi_{\rm H},\Phi)=\g (\Phi_{\rm H}^3/3-M^2\Phi_{\rm H}) + \Phi_{\rm H}\, w_2(\Phi) + w_1(\Phi)$ (with $w_1'(0)=w_2(0)=w_2'(0)=0$).
Concentrating on the heavy field alone, there are two susy preserving vacua with the scalar component being $z_{\rm H}=\pm M$, but there also is an interpolating domain wall solution, preserving half of the supersymmetry. In the limit of a thin wall with infinite tension (finite $M$ and $\g\to\infty$) we then expect the effective superpotentials for the light field in each of the two domains to be different, namely $w_{\rm eff}^\pm(\Phi) \simeq w_1(\Phi) \pm M w_2(\Phi)$ (up to an irrelevant additive constant). In the present paper, however, we will not develop this idea further (apart from a few remarks in an appendix) but instead adopt  a purely ``effective action approach". 

We will consider  ${\cal N}=1$ (rigid) supersymmetric sigma-models with arbitrary K\"ahler potentials 
$K(\Phi^i,\bar\Phi_i)$ and superpotentials $w(\Phi^i)$ for $N$ chiral superfields fields $\Phi^i,\ i=1,\ldots N$ on a domain $\cM$ 
of 3+1 dimensional space-time with a (flat) space-like boundary $\d\cM$. Of course, the presence of the boundary breaks the  super Poincar\'e algebra, with its 4 supersymmetries,  to a 2+1 dimensional one, with 2 supersymmetries. 

After quickly reviewing our conventions for two-component spinors and the usual $3+1$ dimensional superspace, in section 2, we identify the relevant boundary superspace and the generators of the two supersymmetries which remain unbroken by the boundary. Essentially, this corresponds to imposing some reality condition on the supersymmetry parameters $\e$. We will see how, on the boundary, each bulk chiral superfield $\Phi^i$ reduces to a boundary superfield $\f^i$ that consists of two {\it irreducible} boundary superfields (essentially its real and imaginary part). 

Then, in section 3, we consider the susy variation of the standard  four-dimensional sigma-model action $S$. As is well known, the Lagrangian is not invariant but picks up a total derivative which, in the present context, gives rise to a boundary term. We will determine the appropriate minimal (non-supersymmetric) boundary action $\wh S=\int {\rm d}^3 x \ldots$ such that the sum $S+\wh S$ is invariant precisely under the two supersymmetries we identified before as being compatible with the existence of the boundary.\footnote{Henceforth, by ``supersymmetric" we will always mean with respect to the two  supersymmetries preserved by the boundary.} 
 This is achieved without imposing any boundary conditions on the fields. To this ``minimal" action one can further add any susy invariant boundary action of the form
$S_{\rm B}=\int {\rm d}^3 x\, \int {\rm d}^2\t\ B(\f^i, \bar\f_i)$ with real ``boundary potential" $B$. Thus the non-linear sigma-model on a manifold with boundary is characterised by the tripel $(K,B,w)$.

In section 4, we will study the possible supersymmetric boundary conditions on the fields that lead to a well-defined dynamics. Indeed,
boundary conditions on the fields should follow from  a variational principle: as is well known, in order to obtain the Euler-Lagrange field equations by varying some general action $S[\vf^i]$, one has to perform certain partial integrations which generate (further) boundary terms. Boundary conditions must be such that the variation of all boundary terms vanishes.  If one adds an additional boundary action its variation generates an additional boundary term, and thus modifies the resulting boundary conditions. Similarly, if we vary our supersymmetric action $S+\wh S$, we will find the usual field equations in the bulk, as well as  certain boundary terms. The vanishing of the latter can be expressed as {\it algebraic} conditions on the boundary superfields $\f^i$ and $\bar\f_i$ which translate into appropriate Dirichlet and Neumann conditions on the component fields, providing exactly the right amount of boundary conditions. 
We may change these boundary conditions by adding the arbitrary (real) supersymmetric boundary action $S_B$ of the form discussed above.
We will argue that this is the most general boundary term we may add. In particular, terms involving (super)derivatives of the boundary superfield would lead to boundary conditions on the component fields with too many derivatives. 
Any admissible boundary condition then must relate $\f^i$ and $\bar\f_i$ and can be solved, at least locally, as 
$\bar\f_i = G_i(\f^j)$. We will show that these boundary conditions are related to the arbitrary real function $B(\f^i,\bar\f_i)$ by a set of $N$ partial, generally  non-linear differential equation. Note that our whole discussion of boundary conditions is formulated entirely in terms of functions of the boundary {\it superfields} and, hence, invariance under the (boundary) susy algebra is manifest. 

For a single superfield with canonical K\"ahler potential, i.e. the simple Wess-Zumino model, the boundary condition $\bar\f=G(\f)$ would be determined by the ordinary, though still non-linear differential equation:
$
{{\rm d}\over {\rm d} z}\, B\Big(z, G(z)\Big)= 
{i\over 2} \Big( G(z)- z\,G'(z)\Big)$. For this simple model,
we will explicitly work out several non-trivial examples of functions $B$ and  boundary conditions $G$ they determine, and vice versa, including boundary conditions like
$\bar\f = e^{2 i H(\f\, \bar\f)} \f$ with an arbitrary real function $H(x)$, or  $(\bar\f)^2 + \f^2  =1$, or even $\bar\f= \g/ \f$. We will show that {\it any} supersymmetric boundary condition can be obtained by adding an appropriate boundary action $S_B$.

In general, the boundary conditions $G_i$ are solutions of the system of first order partial differential equations containing $B$ and the first derivatives of the K\"ahler potential. Conversely, provided a certain integrability condition (relating the K\"ahler metric and the first derivatives of the $G_i$) 
is satisfied, we prove  that one can obtain a suitable real function $B$ for {\it any} such supersymmetric boundary condition. Note that the superpotential does not enter the discussion of boundary conditions and, hence, it is completely unconstrained by any of the considerations in this paper.

The supersymmetric sigma-model allows for  arbitrary holomorphic field redefinitions $\wt \Phi^i=f^i(\Phi^j)$. It becomes inevitable to consider such holomorphic field redefinitions once one considers general scalar manifolds that are K\"ahler manifolds with non-trivial holomorphic transition functions (like ${\bf CP}^N$). Our whole discussion of supersymmetric boundary conditions is formulated covariantly with respect to such holomorphic field redefinitions. Actually, we will show that there is always a holomorphic field redefinition such that, in terms of the new fields, the boundary conditions simply become $\ov{\wt\f_i}=\wt\f^i$, but this might be at the price of having a complicated K\"ahler potential (and superpotential). This suggests that the boundary conditions can always be understood as defining a real submanifold of the K\"ahler manifold. We will show indeed that the above-mentioned integrability condition is exactly the statement that the submanifold $\zb_i=G_i(z^j)$ is a {\it Lagrangian submanifold}. This is of course analogous to the well-known result in two dimensions that the boundaries of the world-sheet must be mapped to (special) Lagrangian submanifolds of the target manifold.

Consistent boundary conditions should also ensure that the total energy as well as the components of the total momentum tangential to the boundary are conserved. We will show that this is indeed the case with our boundary conditions, provided the total energy includes an appropriate boundary contribution that we identify.

In the remainder of the paper we discuss a few rather obvious generalisations: 
In section 5, we consider the possible coupling of the boundary superfields $\f^i$ to {\it new} (real) superfields $\vf^A$ that only live on the boundary (and have a boundary kinetic term). Using the boundary superspace language, this is straightforward.
Finally, in section 6, we apply the  results of the previous sections to study supersymmetric junctions between different sigma models living on adjacent domains and interacting through a common ``permeable boundary wall": the possible boundary conditions now can mix the boundary superfields of the two sigma models: they become ``matching" conditions. By a folding precedure this situation can be mapped to one with all fields defined on a single domain with a boundary as discussed so far. However, it turns out that a direct analysis is just as simple.
Again, the matching conditions are determined as solutions of a set of partial non-linear differential equations involving the boundary action $S_B$ and the first derivatives of the K\"ahler potentials on both sides of the wall. No condition is imposed on the superpotentials. We give explicit examples of how to match models with different K\"ahler potentials (and superpotentials) in a supersymmetric way.

In Appendix A, we collect a few useful formulae and identities for two-component spinors and superfields. In appendix B, we spell out some details of the proof that a suitable real $B$ exists for every susy boundary condition. Finally, in appendix C, we give some details on the above-mentioned example of how to realize an effective theory having a wall with two different superpotentials on each side, by integrating out a heavy superfield around its domain wall solution. 
%
%
%
\section{Some basic formalism and boundary superspace\label{Algebra}}
\setcounter{equation}{0}
In this section we will identify the relevant boundary superspace, boundary supersymmetry and boundary superfields. First, however, we will recall our conventions which are the same as in ref \cite{abgif} which mostly follow \cite{WB}. 

\subsection{Conventions and bulk superspace\label{convbulk}}

We work in four-dimensional Minkowski space with signature 
$(+---)$.
We use two-component spinors $\p_\a$, $\a=1,2$.  The complex conjugate spinor is denoted with dotted indices: $\pb_\ad \equiv (\p_\a)^*$. Spinor indices are raised and lowered as 
$\p^\a=\e^{\a\b}\p_\b\ , \quad \p_\a=\e_{\a\b}\p^\b \ , \quad
\pb^\ad=\e^{\ad\bd}\pb_\bd\ , \quad \pb_\ad=\e_{\ad\bd}\pb^\bd \ ,
$
where $\e^{12}=-\e^{21}=1$ and $\e_{12}=-\e_{21}=-1$.
The $2\times 2$-matrices $\s^\m$ and $\sb^\m$ are given in terms of 
the standard Pauli matrices $\s_i,\ i=1,2,3$ and the identity matrix by
\be
(\s^\m)_{\a\ad} = ({\bf 1}, -\s_i)_{\a\ad} \quad , \quad
(\sb^\m)^{\ad\a} = \e^{\ad\bd}\e^{\a\b} (\s^\m)_{\b\bd}= 
({\bf 1}, +\s_i)^{\ad\a} \ .
\label{sigmamu}
\ee
When not written explicitly, undotted  indices are always 
contracted from upper left to lower right, while dotted indices are contracted from lower left to upper right, e.g. 
\be
\p\c=\p^\a \c_\a \ , \quad \pb \bar\c = \pb_\ad \bar\c^\ad \ , \quad
\c\s^\m\pb=\c^\a \s^\m_{\a\bd}\pb^\bd \ , \quad
\bar\c\sb^\m\p=\bar\c_\ad\sb^{\m\ad\b}\p_\b \ .
\label{indices}
\ee
Various useful identities between spinor bilinears are given in appendix A.

The $3+1$ superspace has coordinates $x^\m$, as well as an anticommuting (constant) spinor $\t^\a$ and its conjugate $\tb^\ad$. We normalize their integrals as
$\int {\rm d}^2 \t\ \t\t = \int {\rm d}^2 \tb\ \tb\tb =1 \ .
$
While a general superfield is a function of all superspace coordinates, we will be mainly interested in chiral superfields $\Phi$ that obey $\ov{D}_\ad \Phi=0$, cf. app. A, and are functions of 
$y^\m=x^\m+i\t\s^\m\tb$ and $\t$ only:
\be\label{chiralexp}
\Phi(y,\t)=z(y)+\rd\t\p(y)-\t\t f(y)  =  z(x) + \D (x) \ , 
\label{chiral}
\ee
where
\be
\D =  \rd\t\p + i\t\s^\m\tb\, \d_\m z - \t\t f 
-{i\over\rd}\, \t\t\,\d_\m\p \s^\m\tb - {1\over 4}\t\t\,\tb\tb\, \d^2 z \ . 
\label{delta}
\ee
Here and in a more general chiral superfield like $w(\Phi)$ 
the coefficient of $\t\t$ is 
referred to as the $F$-term. Similarly, for a general superfield like
(\ref{gensfield}), the 
coefficient of $\half\t\t\tb\tb$ plus $\half \d_\m\d^\m$ of the lowest 
component (the one without $\t$ and $\tb$) is referred to as the 
$D$-term.

The supersymmetry variation of an arbitrary superfield $S$ reads
\be\label{deltagensfield}
\dd S \equiv (i\e Q+i\eb \overline Q) S \ ,
\quad
Q_\a=-i {\d\over \d \t^\a} - \s^\m_{\a\bd}\tb^\bd {\d\over \d x^\m}
\quad , \quad
\overline Q_\ad=i {\d\over \d \tb^\ad} 
+ \t^\b \s^\m_{\b\ad} {\d\over \d x^\m} \ .
\ee
The supersymmetry generators $Q$ and $\overline Q$ satisfy of course
\be\label{4dsusyalgebra}
\{ Q_\a, \overline Q_\bd\} = 2 \s^\m_{\a\bd} P_\m \ ,
\ee
where $P_\m=-i\d_\m$.
For a chiral superfield, eq. (\ref{deltagensfield}) yields
\be
\dd \Phi 
={\d\over \d y^\m} \left( 2 i\t\s^\m\eb \Phi \right)
+{\d\over \d \t^\a} \left( -\e^\a \Phi\right) \ ,
\label{phisusyvar}
\ee
leading to the component transformations
\be
\dd z=\rd\,\e\p\ , \quad
\dd\p=\rd i\, \d_\m z\, \s^\m \eb - \rd\, f \e \ , 
\quad
\dd f=\rd i\, \d_\m \p \s^\m\eb \ .
\label{fieldvar}
\ee
with the complex conjugate 
expressions reading
\be
\dd \zb=\rd\,\eb\pb \ ,\quad
\dd\pb=-\rd i\, \d_\m \zb\, \e\s^\m  - \rd\, \fb \eb \ , 
\quad
\dd \fb=-\rd i\, \e \s^\m \d_\m \pb  \ .
\label{ccfieldvar}
\ee

\subsection{Boundary superspace and boundary supersymmetry\label{BSBS}}

As is well-known, under infinitesimal Lorentz transformations with parameters $\omega^{\m\n}$ undotted spinors $\p$ and dotted spinors $\pb$ transform as 
\be\label{rotboosts}
\dd_{\rm L} \p_\a= {i\over 2} \omega^{\m\n} (i\s_{\m\n})_\a^{\ \b}\, \p_\b\quad , \quad \dd_{\rm L}\pb^\ad= {i\over 2} \omega^{\m\n}(i\sb_{\m\n})^\ad_{\ \bd}\, \pb^\bd\ ,
\ee
where
$
\s^{\m\n}={1\over 4} \big( \s^\m \sb^\n - \s^\n \sb^\m \big)$ and 
$\sb^{\m\n}={1\over 4} \big( \sb^\m \s^\n-\sb^\n \s^\m \big)$. Somewhat more explicitly, we have 
${i\over 2} \omega^{\m\n} (i\s_{\m\n})={i\over 2}\big( \vec{\a}+i\vec{\n}\big)\cdot \vec{\s}
$ and ${i\over 2} \omega^{\m\n} (i\sb_{\m\n})={i\over 2}\big( \vec{\a}-i\vec{\n}\big)\cdot \vec{\s}\ ,
$
where $\omega^{0j}=\n^j$ (boosts) and $\omega^{kl}=\e^{klj}\a^j$ (rotations) with $j,k,l=1,2,3$.
This shows that $\p$  transforms in the fundamental representation of $SL(2,{\bf C})$ which is the double cover of the proper orthochronous Lorentz group in $3+1$ dimensions, and that $\pb$ transforms in the complex conjugate representation which is different. It is obvious that one cannot impose any relation between $\p$ and $\pb$ without violating the $3+1$ dimensional Lorentz invariance. 

We will be interested in space-times $\cM$ with a flat space-like boundary $\d\cM$, which we take to be the (static) hyperplane $x^n=0$, $n$ being 1, 2 or 3. Specifically, we let $\cM$ be such that $x^n\le 0$ so that the outward normal vector of the boundary $\d\cM$ point towards positive $x^n$. We will often use an index $\mh$ which runs over the three values of $\m$ not being $n$. Transformations that leave the boundary invariant correspond to the $SL(2,{\bf R})$ subgroup of $SL(2,{\bf C})$. (This is most obvious from (\ref{rotboosts}) of one chooses $n=2$.) While the spinor $\p$ transforms irreducibly under $SL(2,{\bf C})$, it contains 2 irreducible components with respect to this $SL(2,{\bf R})$ subgroup.


For a boundary\footnote{
With the appropriate notational modifications, it is trivial to consider any static boundary $\vec{n}\cdot \vec{x}=0$, instead. 
}  
at $x^n=0$, the $2+1$ dimensional Lorentz group is generated by the rotation around the normal vector (parameter $\a^n$) and the two boosts tangential to the boundary (parameters $\n^t$, $t=1, 2, 3$ and $t\ne n$).
Then obviously, $\big( \a^n\s_n + i \n^t \s_t\big) \s_n = \s_n \big( \a^n\s_n - i \n^t\s_t\big)$, and one sees that $\s^n_{\a\bd} \pb^\bd$ transforms exactly as $\p_\a$. It is then consistent to eliminate half of the spinor components by imposing a reality condition
\be\label{realitycond}
\p_\a = - e^{-i\varphi}\, \s^n_{\a\bd} \pb^\bd
\quad \Leftrightarrow \quad
\pb^\ad= e^{i\varphi}\, \sb^{n\ad\b} \p_\b 
\ ,
\ee
where $\varphi$ is an arbitrary phase. A spinor satisfying such a condition then transforms in an irreducible representation of the $SL(2,{\bf R})$ group left unbroken by the boundary at $x^n=0$. 

Similarly, $2+1$ dimensional superspace corresponding to a minimal supersymmetry consists of the $x^\mh$ and a constant 2-component spinor $\t$ obeying the reality constraint (\ref{realitycond}). Here we will call this the boundary superspace ${\cal B}$.
For a single boundary, we can redefine $\t$ by an appropriate phase rotation to give $\varphi$ any desired value. We will find it convenient to fix it as
\be
e^{i\varphi}=-1 \ ,
\ee
so that 
\be\label{boundsuperspace}
{\cal B}=\{ x^\m, \t, \tb \ \vert \ x^n=0\ ,\ \tb^\ad= -\sb^{n\ad\b} \t_\b \} \ .
\ee
It will be useful to have at hand various equivalent ways to write the constraint on $\t$:
\be\label{thetareal}
\tb^\ad= -\sb^{n\ad\b} \t_\b 
\quad\Leftrightarrow \quad
\tb_\ad=\t^\b \s^n_{\b\ad}
\quad\Leftrightarrow \quad
\t_\a=\s^n_{\a\bd}\tb^\bd
\quad\Leftrightarrow \quad
\t^\a=-\tb_\bd \sb^{n\bd\a}
\ee 
We also note that this implies
\be\label{tstb}
\t\s^\m\tb= \dd^\m_n\ \t\t \ .
\ee
A boundary superfield $\wh \f$ is a field defined on ${\cal B}$.
It has the expansion 
\be\label{genbsf}
\wh \f(x^\mh,\t)=\z(x^\mh) + \rd \t \chi(x^\mh) - \t\t g(x^\mh) \ .
\ee
In  general, the component fields $\xi, \chi$ and $g$ may be real or complex.

Since supersymmetry corresponds to translations in superspace, the boundary supersymmetry must have parameters $\e$ satisfying the same constaint as $\t$:
\be\label{epsreality}
\eb^\ad= - \sb^{n\ad\b} \e_\b 
\quad\Leftrightarrow \quad
\eb_\ad=\e^\b \s^n_{\b\ad}
\quad\Leftrightarrow \quad
\e_\a=\s^n_{\a\bd}\eb^\bd
\quad\Leftrightarrow \quad
\e^\a=-\eb_\bd \sb^{n\bd\a}\ .
\ee
For any boundary superfield, the bulk supersymmetry then reduces to
\be\label{boundsusyvar}
\dd_{\rm susy} \wh\f= i(\e Q + \eb \overline Q)\wh\f
=i (\e^\a Q_\a + \e^\b \s^n_{\b\ad} \overline Q^\ad)\wh\f
= i \e^\a \wt Q_\a \wh\f\ ,
\ee
with $\wt Q_\a=Q_\a + \s^n_{\a\bd} \overline Q^\bd$ being the boundary supersymmetry generator. Explicitly, we find\footnote{
One has to be careful when computing $\d/\d\t$ and $\d/\d\tb$ on the boundary superfields with $\t$ and $\tb$ constrained by $\tb^\ad= - \sb_n^{\ad\b} \t_\b$. One should first  change variables to $\t_{(1)}^\a={1\over 2} ( \t^\a + \tb_\bd \sb^{n\bd\a})$ and $\t_{(2)}^\a={1\over 2} ( \t^\a - \tb_\bd \sb^{n\bd\a})$. Then  any boundary superfield only depends on $\t_{(1)}$. One has ${\d\over\d\t^\a} \f(\t_{(1)})= {1\over 2} {\d\over\d\t_{(1)}^\a} \f(\t_{(1)})$, etc, resulting in an extra factor ${1\over 2}$.
}
\be\label{susygenerators}
\wt Q_\a\equiv Q_\a + \s^n_{\a\bd} \overline Q^\bd = -i {\d\over \d\t^\a} + 2 \,\wh\g^\mh_{\a\b} \t^\b {\d\over \d x^\mh} \ ,
\quad {\rm with} \quad 
\wh\g^\mh_{\a\b}= \wh\g^\mh_{\b\a}=- \s^\mh_{\a\gd} \e^{\gd\dot\dd}\s^n_{\b\dot\dd} ,
\ee
where the $\wh\g^\mh$ obey the 2+1 dimensional Clifford algebra.
It is not difficult to see that these $\wt Q_\a$ then satisfy the 2+1 dimensional supersymmetry algebra $\{\wt Q_\a, \wt Q_\b\}=2 \wh\g^\mh_{\a\b}\, P_\mh$ as appropriate on the boundary.

We will be mainly interested in the boundary superfields obtained by restricting the bulk (anti) chiral superfields $\Phi$ and $\bar\Phi\equiv\Phi^\dag$ to the boundary superspace ${\cal B}$. Using (\ref{tstb}), the restriction of  $y^\m=x^\m +i\t\s^\m\tb$ yields $y^\mh\vert_{\cal B}=x^\mh$ and $y^n\vert_{\cal B}= i\t\t$, so that one gets (cf.~(\ref{chiralexp})):
\ba\label{chiralbsf}
\f= \Phi\vert_{\cal B} &=& z + \rd \t\p - \t\t (f-i\d_n z) \ ,
\nonumber\\
\bar\f = \bar\Phi\vert_{\cal B} &=& \zb + \rd \t\s^n\pb - \t\t (\fb+i\d_n \zb) \ ,
\ea
where it is understood that the component fields have arguments $x^\mh$, while $x^n=0$. Note that $\bar\f\equiv \f^\dag$. One can now apply (\ref{boundsusyvar}) on $\f$ and on $\bar\f$ to obtain the variations of the component fields under the boundary supersymmetry (with $\e$ satisfying (\ref{epsreality})):
\ba\label{boundsusyfieldtransf}
\dd z=\rd \e\p \quad &,& \quad \dd \zb=\rd\,\e\s^n\pb
\nonumber\\
\dd\p=-\rd i\, \d_\mh z \s^\mh \sb^n \e -\rd (f-i\d_n z)\,\e
\quad &,& \quad 
\dd\pb=-\rd i\,\d_\mh\zb \e \s^\mh -\rd (\fb+i\d_n\zb)\, \e\s^n
\nonumber\\
\dd (f-i\d_n z) = -\rd i\, \d_\mh\p \s^\mh \sb^n \e 
\quad &,&\quad
\dd(\fb+i\d_n\zb) =-\rd i\, \e\s^\mh \d_\mh \pb \ .
\ea
Of course, the same variations are also obtained directly from the bulk variations (\ref{fieldvar}) and (\ref{ccfieldvar}) upon imposing (\ref{epsreality}) on the susy parameter $\e$.

Let us summarize: In this section, based on symmetry algebra considerations, we have identified the two supersymmetries that remain unbroken by the boundary, cf.~(\ref{boundsusyvar}). This corresponds to imposing the reality constraint (\ref{epsreality}) on the susy parameters $\e$. This unbroken supersymmetry is most conveniently described using the boundary superspace ${\cal B}$ defined in (\ref{boundsuperspace}). When restricted to ${\cal B}$, every bulk chiral superfield $\Phi$ gives rise to a boundary superfield $\f$. So far, we have not imposed any boundary conditions on these fields (i.e. on their component fields). The question of boundary conditions will only be considered later-on in sect. 4, but it is clear that any boundary condition formulated in terms of the boundary superfields will automatically preserve the unbroken supersymmetry (\ref{boundsusyvar}).

We should note that the present discussion immediately generalizes to two {\it parallel} boundaries, say at $x^n=-a$ and $x^n=0$. Since   $\e$ is a global parameter, the supersymmetry preserved by both boundaries must be the same, the generators being given by (\ref{susygenerators}). In particular, the arbitrary phase $\vf$ must be chosen identically for both boundary superspaces.

%
%
\section{Susy invariance of the sigma-model bulk plus boundary action \label{Action}}
\setcounter{equation}{0}

In this section we identify the appropriate boundary action to be added to the usual bulk action of the non-linear sigma-model in order to achieve invariance under the two supersymmetries that remain unbroken by the boundary. We will {\it not} impose any boundary conditions.

\subsection{The standard action\label{standac}}

We want to consider a general ${\cal N}=1$ susy non-linear
sigma-model for $N$ chiral fields $\Phi^i$ and their hermitian conjugates $\bar\Phi_i$. The standard bulk action can be written in terms of (bulk) superspace integrals as
\be
S_{\s-\rm model}^{(1)}=\intm \left( \intttb K(\Phi^i, \bar\Phi_j) 
+\intt w(\Phi^i) + \inttb \bar{w}(\bar\Phi_i) \right)
\equiv S_K^{(1)} + S_w^{(1)} + S_{\bar{w}}^{(1)} \ ,
\label{sigaction}
\ee
or alternatively  as
\be
S_{\s-\rm model}^{(2)}=\intm \left(\half \left[ K(\Phi^i, \bar\Phi_j) \right]_D
+\left[ w(\Phi^i) \right]_F 
+ \left[\bar{w}(\bar\Phi_i) \right]_{\bar{F}} \right)
\equiv S_K^{(2)} + S_w^{(2)} + S_{\bar{w}}^{(2)} \ .
\label{S2}
\ee
Here $[\ldots]_D$ and $[\ldots]_F$ refer to picking out the 
$D$-terms and $F$-terms 
(resp. $\bar F$-terms). 
$K(z,\zb)$ is the real K\"ahler potential: $[K(z^i,\zb_j)]^*=K(z^i,\zb_j)$ and $w(z^i)$ the holomorphic superpotential. Expanding $K$ and $w$ and doing the superspace integrals, resp. picking out the $D$ and $F$ terms, leads to the standard bulk action for the component fields which we will give below. This involves various derivatives of the potentials, for which we use the 
standard notation
\be\label{Kinot}
K_i={\d K(z,\zb)\over \d z^i}\ , 
\quad K^j={\d K(z,\zb)\over \d\zb_j}\ ,\ 
K_i^j= {\d^2 K(z,\zb)\over \d z^i \d\zb_j}    \ , \ 
w_i={\d w(z)\over \d z^i}\ , \ 
w^j={\d \bar{w} \over \d\zb_j} \ , \ 
{\rm etc...}
\label{kikj}
\ee
They obey certain reality conditions, e.g.
\be
(K^i_j)^*=K^j_i \quad , \quad (K^i_{jk})^*=K_i^{jk} 
\quad , \quad (w^j)^*=w_j\ .
\label{herm}
\ee

Both Lagrangians (\ref{sigaction}) and 
(\ref{S2}) only differ by a total derivative and, in the absence of boundaries, lead to identical actions. Of course, in the presence of a boundary the actions differ by a boundary term. Since later-on we will anyhow have to add a boundary term, one might be  tempted to start with either action. However, $S_w^{(1)}$ gives a boundary term that still depends on $\tb$ -- which is clearly unwanted. Thus the correct action to begin with is $S^{(2)}$.
The component expansions are
\ba\label{Dtermexpl}
\hskip-1.5cm
\half \left[ K(\Phi^i, \bar\Phi_j) \right]_D
&=&
K_i^j \left( f^i\fb_j+\d_\m z^i\d^\m\zb_j
-{i\over 2}\p^i\s^\m\d_\m\pb_j + {i\over 2}\d_\m\p^i\s^\m\pb_j \right)
\nonumber\\
&&+\ {1\over 2}K^k_{ij} \left( i\, \p^i\s^\m\pb_k\d_\m z^j 
+\p^i\p^j\fb_k\right) +\ h.c.\ 
+\ {1\over 4} K_{ij}^{kl} \p^i\p^j\pb_k\pb_l \ ,
\ea
and
\be
\left[ w(\Phi^i) \right]_F = - w_i f^i(x) -\half w_{ij} \p^i(x)\p^j(x) \ .
\label{wcomp}
\ee
As is well-known, after adding the $D$ and $F$ terms, the fields $f^i$ and $\fb_j$ only appear algebraically  and at most bilinearly and, hence, could be eliminated. In this section, we will not do so, however, and stay completely off-shell.%
%
%
\subsection{Boundary terms from the susy variation of the bulk action $S^{(2)}$\label{btfsv}}
%

As recalled in (\ref{Fsusyvar}) and (\ref{phisusyvar}), the susy 
variations of superfields are total derivatives in superspace. 
Hence, if $\cM$ has no boundary, $S$ is invariant under 
supersymmetry. At present, however,
$\cM$ has a boundary and $\dd S$ will pick up boundary terms. To begin with, we determine these boundary terms for an arbitrary supersymmetry transformation, i.e. with no restriction on the parameter $\e$.

First, consider the $F$-terms: it is easiest to look directly 
at the component expression (\ref{wcomp}) and use (\ref{fieldvar}). 
One gets
\be
\dd \left[ w(\Phi^i) \right]_F 
= -i \rd \d_\m \left( w_j \p^j \s^\m \eb \right)
\label{deltaw}
\ee
where we used $w_{ijk}\, \p^i\p^j\ \e\p^k=0$. Hence
\be
\dd S_w^{(2)} = 
\dd\intm \left[ w(\Phi^i) \right]_F = -i\rd \intdm w_j\, \p^j\s^n \eb \ .
\label{wvar2}
\ee
Taking the complex conjugate yields similarly (recall $w_j^*\equiv w^j$)
\be
\dd S_{\bar{w}}^{(2)}
= i\rd \intdm w^j\, \e \s^n\pb_j 
=-i\rd \intdm w^j\, \pb_j\sb^n\e  \ .
\label{wbarvar}
\ee

To compute the susy variation of the $D$-term $S_K^{(2)}$ is a bit more involved. After a somewhat lengthy computation\footnote{
We found it easiest to first compute the susy variation of $S_K^{(1)}$ and then add the susy variation of the boundary term $S_K^{(2)}-S_K^{(1)}$. Apart from (\ref{thetaidentities}), a useful identity is
$K_i\d_\m\p^i+K_{ij}\d_\m z^j\p^i-K_i^j\d_\m\zb_j\p^i$
$=\d_\m(K_i\p^i) - 2 K_i^j\d_\m\zb_j\p^i$. One also uses the fact that one can freely do partial integrations on the boundary since the latter cannot have a boundary itself. Hence
 $\intdm\d_\m(K_i\p^i) \s^\m\sb^n\e
= - \intdm\d_n(K_i\p^i) \e$.
} 
we find
\be
\dd S_K^{(2)}
={1\over \rd} \intdm \Bigg\{  
i\left(  K_j^i f^j + \half K^i_{jk}\p^j\p^k\right) \pb_i \sb^n\e 
+  K_i^j\d_\m\zb_j\p^i  \s^\m\sb^n\e \Bigg\} + h.c. 
\label{Kvar2}
\ee
Finally, putting all the pieces together, we get the variation of the bulk action $S$ under an arbitrary supersymmetry with (unrestricted) parameter $\e$\ :
\be
\dd S^{(2)}={1\over \rd} \intdm\left\{  
i\left(K_j^i f^j +\half K^i_{jk}\p^j\p^k -2 w^i\right) \pb_i\sb^n\e 
+K_j^i\d_\m\zb_i\p^j \s^\m\sb^n\e \right\} + h.c. 
\label{Svar2}
\ee
%
%
%
\subsection{Adding an appropriate boundary action $\wh S$\label{Shatsect}}
%
Obviously, this variation $\dd S^{(2)}$ of the bulk action is non-vanishing.
As discussed above, in the presence of a boundary,  we can only hope to preserve half of the initial $3+1$ dimensional supersymmetry. In particular, we have determined the conditions (\ref{epsreality}) to be imposed on $\e$ in order to get a consistent $2+1$ dimensional super-Poincar\'e algebra.
Imposing these conditions on $\e$ (we write $\eb=\e \s^n$) and using the identities of appendix A, we can rewrite the variation (\ref{Svar2}) in the following somewhat simpler form:
\ba\label{deltaSeb=e}
\hskip-1.cm
\dd S^{(2)}\Big\vert_{\eb=\e \s^n}={1\over \rd} \intdm\Big\{ &&\hskip-7.mm
-K_j^i\d_\m z^j\, \e\s^\m\pb_i - K^i_j \d_\m\zb_i\, \p^j\s^\m\eb
+2i\, w^i\,\eb\pb_i -2i\, w_i\, \e\p^i
\nonumber\\
&&\hskip-7.mm
-i\big( K^i_j f^j+\half K^i_{jk}\p^j\p^k\big)\, \eb\pb_i
+i\big( K_i^j \fb_j+\half K_i^{jk}\pb_j\pb_k\big)\, \e\p^i
\Big\}\Big\vert_{\eb=\e \s^n}\ .
\ea
This is still non-vanishing. At this point we have two possibilities to recover invariance under the supersymmetry restricted by $\eb=\e\s^n$. Since (\ref{deltaSeb=e}) only involves the fields on the boundary, one could try to impose boundary conditions on these fields. However, this turns out not to be necessary. Instead, we observe that, as in any field theory, we could have started with a Lagrangian that differs from ours by a total derivative, i.e. we have the freedom to add any boundary term to our action. Such a boundary term should, of course, preserve the symmetries of our theory. At present, we will try to find a ``minimal", $2+1$-Poincar\'e invariant boundary action $\wh S$ such that its susy variation (subject to $\eb=\e\s^n$) precisely cancels (\ref{deltaSeb=e}). By ``minimal" we mean that it should only involve the fields, K\"ahler potential and superpotential (and their derivatives) already present in $S^{(2)}$, and no additional fields or functions. Later-on we will also add additional non-minimal boundary actions $S_B$ that are susy invariant by themselves and, hence, can be written in terms of the boundary superfields. On the other hand, the non-invariant action $\wh S$ {\it cannot} be written using the boundary superfields only, and we have to find its expression in component fields. 

To figure out the appropriate boundary action $\wh S$ we first concentrate on the term in (\ref{deltaSeb=e}) that contains $K^i_{jk}\p^j\p^k\, \eb\pb_i$. Observe that, since each $\p^i$ has only two anticommuting components, we have $K_{jkl} \p^j\p^k\p^l=0$. Then one finds
\be\label{Kpsipsi}
{1\over \rd} \dd \big( K_{jk}\p^j\p^k\big) =
K^l_{jk} \p^j\p^k\, \eb\pb_l 
+2K_{jk} \p^j(i\d_\m z^k \s^\m \eb-f^k\e) \ .
\ee
Furthermore,
\be\label{Kfi}
{1\over \rd} \dd \big( K_j f^j\big) = K_{jl} f^j\, \e\p^j 
+ K_j^l f^j\, \eb\pb_l + i K_j \d_\m \p^j\s^\m \eb \ .
\ee
Combining both equations and using $K_{jk}\d_\m z^k=\d_\m K_j - K_j^i \d_\m\zb_i$, as well as the complex conjugate relations, one gets
\ba
&&\hskip-1.cm\dd \Big( -{i\over 2} K_j f^j -{i\over 4}  K_{jk}\p^j\p^k +{i\over 2} K^j \fb_j + {i\over 4} K^{jk} \pb_j \pb_k -i w + i \bar{w}\Big)
\nonumber\\
&&={1\over \rd}  \d_\m \Big( K_j\, \p^j\s^\m\eb + K^j\, \e\s^\m\pb_j\Big) 
\nonumber\\
&&-K_j^i\d_\m z^j\, \e\s^\m\pb_i - K^i_j \d_\m\zb_i\, \p^j\s^\m\eb
-i\big( K^i_j f^j+\half K^i_{jk}\p^j\p^k\big)\, \eb\pb_i
+i\big( K_i^j \fb_j+\half K_i^{jk}\pb_j\pb_k\big)\, \e\p^i
\nonumber\\
&&-2i\, w_i\, \e\p^i +2i\, w^i\,\eb\pb_i \ .
\ea
Except for the first one, the terms on the right hand side are exactly what we need to cancel the susy variation (\ref{deltaSeb=e}) of the bulk action.
To deal with the first term, note again that the boundary itself does not have a boundary and hence 
\be
\intdm \d_\m \Big( K_j\, \p^j\s^\m\eb + K^j\, \e\s^\m\pb_j\Big) 
= \intdm \d_n \Big( K_j\, \p^j\s^n\eb + K^j\, \e\s^n\pb_j\Big) \ .
\ee
If we now restrict the susy transformation parameter $\e$ to obey (\ref{epsreality}), we get
\be
{1\over \rd}\intdmh \d_\m \Big( K_j\, \p^j\s^\m\eb + K^j\, \e\s^\m\pb_j\Big)\Big\vert_{\eb=\e \s^n}={1\over \rd} \intdmh \d_n \Big( K_j\, \p^j\e + K^j\, \eb\pb_j\Big)=\dd \intdmh\half \d_n K \ .
\ee
Combining everything, we finally find
\be
\dd \big(S^{(2)}+\wh S)\Big\vert_{\eb=\e \s^n}=0 \ ,
\ee
with our (minimal) boundary action $\wh S$ being
\be\label{Shat}
\wh S={1\over 2} \intdm \left\{  \d_n K 
+i \Big( K_j f^j +\half K_{jk}\p^j\p^k \Big)
-i \Big(K^j \fb_j + \half K^{jk} \pb_j \pb_k \Big)
+2i w - 2i \bar{w} \right\} \ .
\ee
For later reference, we note that we can also rewrite this as
\be\label{Shat2}
\wh S=- \intdm\, \Im \left\{  
K_j \big(f^j-i\d_n z^j\big) +\half K_{jk}\p^j\p^k 
+2 w \right\}  \ .
\ee
We have thus achieved to identify a boundary action $\wh S$ to be added to the bulk action $S^{(2)}$ such that the sum is invariant under the two supersymmetries preserved by the boundary, i.e. with $\eb=\e \s^n$. We conclude that the correct action for the supersymmetric $\s$-model in the presence of a boundary is not just $S^{(2)}$ but 
\be\label{Stilde}
S=S^{(2)}+\wh S \ .
\ee

Note again that susy invariance of $S=S^{(2)}+\wh S$ is achieved without imposing any boundary conditions on the fields. On the other hand, it is now also clear that if one had insisted not to add $\wh S$, one would have had to impose boundary conditions such that $\wh S$ vanishes. One sees from (\ref{Shat2}) that such boundary condition would have to be such that $K_j \big(f^j-i\d_n z^j\big) +\half K_{jk}\p^j\p^k 
+2 w$ is real, i.e. typically (up to field redefinitions) $z^j$ real, $f^j-i\d_n z^j$ real, $\p^j$ satisfying the same reality conditions as $\e$ or $\t$, and appropriate reality conditions on $K_j$, $K_{jk}$ and on the superpotential $w$. We will find such boundary conditions {\it on the fields} as a special case in the next section when studying the dynamics. However, we will not get any reality condition {\it on the superpotential}.

\subsection{Additional susy invariant boundary actions $S_B$\label{addsuiba}}

Instead of just adding the minimal $\wh S$ to the bulk action $S^{(2)}$, one can also add further boundary actions that are by themselves invariant under the supersymmetries satisfying $\eb=\e\s^n$. Such invariant boundary actions are easy to construct. Indeed, as discussed in section \ref{BSBS}, the boundary superspace integral of any function of boundary superfields is invariant. Thus, we can add a boundary action
\be\label{SBdef}
S_B=\intdm \intt B(\f^l,\bar\f_k) \ ,
\ee
with an arbitrary real function $B(z^l,\zb_k)$.
More general invariant boundary actions could also involve boundary superderivatives, as would be appropriate to describe new degrees of freedom that have their own $2+1$ dimensional propagator.
To write the component form of (\ref{SBdef}), we denote, in analogy with (\ref{Kinot})
\be\label{Binot}
B_i={\d B(z^l,\zb_k)\over \d z^i} \ , \quad
B^j={\d B(z^l,\zb_k)\over \d \zb_j} \ , \quad {\rm etc.} 
\ee
Expanding (\ref{SBdef}) and using (\ref{chiralbsf}) then gives
\be\label{SBcompexp}
S_B=-\intdm \Big[
B_j (f^j-i\d_n z^j) + B^j (\fb_j+i\d_n\zb_j) 
+B_i^j \p^i\s^n\pb_j +\half B_{ij} \p^i\p^j +\half B^{ij} \pb_i\pb_j 
\ \Big] \ .
\ee
Thus the supersymmetric sigma-model on a manifold with boundary is characterized by the tripel $(K,B,w)$.

\section{Boundary conditions from stationarity of the action\label{bcandeom}}
\setcounter{equation}{0}

As is well-known, the (classical) action of a field theory encodes the field equations and boundary conditions. Indeed, in order to obtain the field equations  upon varying the action,  one has to perform some partial integrations. This generate boundary terms. Stationarity of the action requires the vanishing of the bulk terms, which yields the Euler-Lagrange field equations, {\it and} of the boundary terms, which yields boundary conditions. In general, one can add a boundary action (respecting the various symmetries) which, upon variation, yields extra boundary terms and, hence, modifies the boundary conditions. These boundary conditions should also ensure the conservation of the (appropriately modified) total energy and the components of the total momentum that are parallel to the boundary.

In this section, after briefly looking at the simple example of a single scalar field, we will carry out this program for our supersymmetric non-linear sigma-model action $S=S^{(2)}+\wh S$. (Henceforth, ``supersymmetric" means invariant under the two supersymmetries preserved by the boundary, i.e. satisfying $\eb=\e\s^n$.) We will identify the boundary terms generated upon varying the fields and express them entirely in terms of the boundary superfields. This leads to simple supersymmetric boundary conditions. Adding a general supersymmetric boundary action $S_B$ of the form (\ref{SBdef}), we can generate extra boundary terms and obtain general supersymmetric boundary conditions.  More precisely, we will show that for {\it every} $S_B$, consistent supersymmetric boundary conditions are obtained as solutions of a set of partial differential equations involving $B$ and the first derivatives of the K\"ahler potential. (For a single chiral superfield one just has an ordinary differential equation.) Conversely, every supersymmetric boundary condition is shown to determine (modulo an integrability condition) an appropriate (non-unique) boundary action $S_B$. This integrability condition has a nice geometric interpretation: it states that, on the K\"ahler manifold, the boundary conditions must define a Lagrangian submanifold
We provide many examples of boundary actions and boundary conditions they determine, and vice versa. The whole discussion of boundary conditions will be in terms of boundary superfields and, hence, will be {\it manifestly supersymmetric}. It will also be {\it manifestly covariant} with respect to arbitrary holomorphic field redefinitions of the non-linear sigma model. We will also show that the total energy (possibly including a boundary energy), as well as the two tangential components of the total momentum are indeed conserved, once the bulk equations of motion and the boundary conditions are imposed.

A remark on notation: in this section we will consider arbitrary field variations, denoted again by $\dd z$, $\dd\p$ etc. These are {\it not} meant to be the susy variations. Similarly, $\dd S$ will denote the variation of an action under these field variations.

\subsection{A warm-up exercice: a single scalar field\label{scalarex}}

It is instructive to first recall the case of a single scalar field with bulk action
\be\label{scalarbulk}
S_{\rm bulk}=\intm\left[ \half K(\vf)\,\d_\m\vf\d^\m\vf -V(\vf)\right] \ .
\ee
Upon varying the field $\vf$ one gets
\be\label{deltaSbulk}
\dd S_{\rm bulk}=\intm [\ldots] \dd\vf - \intdm K(\vf) \d_n\vf \dd\vf \ ,
\ee
where $[\ldots]=0$ is the Euler-Lagrange field equation for $\vf$.
Requiring $\dd S_{\rm bulk}=0$ then implies this field equation {\it and} a boundary condition on $\vf$ such that the boundary term vanishes. Clearly, the latter is either $\d_n\vf\vert_{\d\cM}=0$ (Neumann) or $\dd\vf\vert_{\d\cM}=0$, i.e. $\vf\vert_{\d\cM}=const$ (Dirichlet). Let us insist that we have two possibilities and the boundary conditions are not determined uniquely.

We can add an extra boundary action (which can depend on $\vf$ and $\d_n\vf$ since both are $2+1$ dimensional scalar fields)
\be\label{scalarbounaction}
S_{\rm bound}=\intdm\, b(\vf,\d_n\vf)
\quad \Rightarrow\quad
\dd S_{\rm bound}=\intdm \Big( \d_1 b(\vf,\d_n\vf)\ \dd\vf + 
\d_2 b(\vf,\d_n\vf)\ \d_n\dd\vf\Big) \ ,
\ee
where $\d_1 b$ and $\d_2 b$ denote the derivatives of $b$ with respect to its first and second arguments. Vanishing of the total boundary term now requires (on $\d\cM$)
\be\label{brel}
\Big( \d_1 b(\vf,\d_n\vf)- K(\vf) \d_n\vf \Big)\ \dd\vf + 
\d_2 b(\vf,\d_n\vf)\ \d_n\dd\vf =0\ .
\ee
This can only be satisfied if there is a relation between $\vf$ and $\d_n \vf$ on $\d\cM$, i.e. a boundary condition.

Let us first assume that the boundary condition does not involve $\d_n\vf$. It is then of the form $h(\vf)=0$ which is generically solved with $\vf$ equal to one of the roots of $h$. Hence this is a Dirichlet condition and implies $\dd\vf=0$. Relation (\ref{brel}) then implies $\d_2 b(\vf,\d_n\vf)=0$, i.e. $S_{\rm bound}=\intdm\wt b(\vf)$.

If the boundary condition does involve $\d_n\vf$, generically we can solve it as
$
\d_n\vf= g(\vf)$ which implies $\dd\d_n\vf=g'(\vf)\dd\vf \ .
$
For $g=0$ this is a Neumann condition, while for general $g$ it could be called a mixed boundary condition.
Equation (\ref{brel}) then gives
$\Big( \d_1 b(\vf,g(\vf))- K(\vf) g(\vf) \Big)\dd\vf + 
\d_2 b(\vf,g(\vf))\ g'(\vf)\dd\vf=0$, which can be rewritten as\footnote{
For $g\ne 0$ one cannot have a Dirichlet condition and, hence, we can regard $\dd\vf$ as the unconstrained variation.
}
\be\label{bgdiffeq}
{{\rm d} \over {\rm d}\vf}\ b\big(\vf,g(\vf)\big) = K(\vf)\,g(\vf) \ .
\ee
We see that for given function $b$, i.e. for given boundary action $S_B$, the boundary condition, i.e. the function $g$, is determined as a solution of this generally non-linear ordinary differential equation.\footnote{
As a simple example, consider $b(\vf,\d_n\vf)= {1\over \g} \vf\, \d_n\vf$ and $K=1$. Then the differential equation (\ref{bgdiffeq}) is solved by $g(\vf)\sim \vf^{\g-1}$, i.e. a boundary condition $\d_n\vf\sim \vf^{\g-1}$. Conversely, for given $g$ it is not difficult to find an appropriate function $b(x,y)$ such that $g$ is a solution of (\ref{bgdiffeq}): any $b_\a$ with $b_\a(x,y)=\left({y\over g(x)}\right)^\a \int^x {\rm d}x' K(x') g(x')$ does the job. Let us note that for $\a=0$ the same $b$ allows for both, $\d_n \vf=g(\vf)$ and for Dirichlet conditions.
}
If one had several scalar fields, one would similarly get a set of non-linear partial differential equations. The non-linearity implies again that, in general, there are several solutions.

One can ask whether one could add a boundary action involving higher, e.g. second order normal derivatives like $\intdm a(\vf,\d_n\vf,\d_n^2\vf)$ to obtain a boundary condition like $\d_n^2\vf=f(\vf,\d_n\vf)$. Going through a similar analysis as before,  instead of (\ref{bgdiffeq}) one now gets two partial differential equations: ${\d\over\d x} a\big(x,y,f(x,y)\big) = K(x)\, y$ and ${\d\over\d y} a\big(x,y,f(x,y)\big)=0$. These two equations are incompatible, showing that boundary conditions involving second-order normal derivatives cannot follow from a variational principle. Obviously, the same conclusion applies for even higher-order normal derivatives. 


Another, maybe more physical aspect of boundary conditions is that they should ensure conservation of energy and momentum. Of course, the presence of the boundary breaks translational invariance in the $x^n$-direction and we only expect $P^\mh$, i.e. the total energy and the total tangential momenta to be conserved\footnote{
Recall that $\mh\ne n$}. 
We will see that the boundary conditions determined by $b(\vf, \d_n\vf)$ are exactly such that the naive momenta $P^\mh,\ \mh\ne 0$ are conserved and that $P^0$
is conserved once we add the ``boundary energy" $\int{\rm d}^2 x\, b$.

First note that translational invariance in all four space-time directions of the bulk Lagrangian ${\cal L}={1\over 2} K(\vf) \d_\m\vf\d^\m\vf -V(\vf)$ and the Euler-Lagrange field equations imply
\be\label{tmn1}
\d_\m T^\m_{\ \, \n}=0 \quad , \quad T^\m_{\ \, \n} = {\d{\cal L}\over \d\, \d_\m\vf} \d_\n\vf - \dd^\m_\n\, {\cal L} 
\ee
everywhere in $\cM$, even on the boundary $\d\cM$. In particular, we have $T_{n\nh}= K(\vf)\, \d_n\vf\, \d_\nh\vf$. Similarly, translational invariance in the three boundary directions $x^\mh$ of the ``boundary Lagrangian" $b(\vf,\d_n\vf)$ (i.e. no explicit dependence on $x^\mh$), together with the boundary condition $\d_n\vf=g(\vf)$ imply
$\d_\nh\, b(\vf,\d_n\vf) = \d_\nh\, b(\vf,g(\vf))
= \d_\nh\vf \, {{\rm d}\over {\rm d}\vf} b(\vf,g(\vf))
=\d_\nh\vf\, K(\vf) g(\vf)  \ ,
$
where we used (\ref{bgdiffeq}) in the last step. This, together with the boundary condition, shows that on the boundary $T_{n\nh}$ is a total derivative:
\be\label{tmn3}
T_{n\nh}= K(\vf)\, g(\vf) \d_\nh\vf = \d_\nh\, b \qquad {\rm on} \ \d\cM \ .
\ee
We now define the total energy and momenta $P^\nh$ as
\be\label{tmn4}
P^\nh=\int_{x^n\le 0} {\rm d}^3 x\, T^{0\nh} - g^{0\nh} \int_{x^n=0}{\rm d}^2 x\, b \ .
\ee
Clearly, $-b$ plays the r\^ole of a ``boundary potential" and must contribute to the total energy. Then, using $\d_\m T^{\m\nh}=0$, we have
\be\label{tmn5}
{{\rm d}\over {\rm d} t} P^\nh = \int_{x^n=0} {\rm d}^2 x\, \left( -T^{n\nh} - g^{0\nh} \d_0 b\right) 
=\int_{x^n=0} {\rm d}^2 x\, \left( + \d^\nh b - g^{0\nh}\d_0 b\right) =0 \ .
\ee
Indeed, for $\nh=0$ the integrand vanishes, while for $\nh\ne 0$ the integrand is a total derivative in one of the two tangential directions so that the integral vanishes, too. 

\vskip2.mm
\centerline{***}

In the remainder of this section, we will apply similar arguments to the supersymmetric sigma-model starting from our supersymmetric action $S=S^{(2)}+\wh S$. 

\subsection{Boundary terms from varying $S+S_B$\label{btvsh}}

To begin with, we determine the boundary terms that result from varying the fields in $S=S^{(2)}+\wh S$ and doing the partial integrations. Of course, since $\wh S$ is a boundary action it only contributes to the boundary term and not to the field equations. Schematically, calling the fields $\xi^a$, we have
\be
\dd S^{(2)}= \intm ({\rm eom})_a\, \dd \xi^a 
+ \intdm\, \Sigma^{(2)}
\quad , \quad
\dd \wh S= \intdm\, \wh\Sigma \ ,
\ee
where $({\rm eom})_a$ is the Euler-Lagrange field equation for the $a^{\rm th}$ field $\xi^a$, and $\intdm \big( \Sigma^{(2)}+\wh \Sigma\big)$ is the boundary term we want to determine.  Since the total action $S$ is supersymmetric, its variation $\dd  S$ necessarily also is supersymmetric. However, the bulk and the boundary terms  have no reason to be separately supersymmetric, and indeed they will turn out not to be. Nevertheless, if the field equations are satisfied everywhere in the bulk (and hence also on the boundary) the bulk term vanishes and the boundary term must be supersymmetric by itself. In fact, we will find that we do not need to impose all field equations, it will be enough to impose the (algebraic) field equations for the auxiliary fields in order to make the boundary term supersymmetric. Imposing a bulk field equations on the boundary term certainly does not constitute any restriction when requiring that the total variation of the action vanishes: the bulk and boundary terms both have to vanish. Vanishing of the first gives the field equations of motion and we can certainly use them at the boundary to simplify the boundary term.
 
The boundary term $\Sigma^{(2)}$ arise from the partial integrations, and to determine it, it is enough to look at the $D$-term (\ref{Dtermexpl}). We find
\be\label{Sigma}
\Sigma^{(2)}= \left\{-\dd z^i \left( K_i^j\d_n\zb_j-{i\over 2} K^k_{ij}\p^j\s^n\pb_k\right) +{i\over 2} K_i^j \dd\p^i \s^n\pb_j \right\} + h.c. \ .
\ee
Next, from (\ref{Shat2}) we get, after slightly rearranging the terms
\ba\label{Sigmahat}
\wh\Sigma&=&\Bigg\{-{i\over 2} K_i^j\, \dd z^i (\fb_j+i\d_n\zb_j) 
+{i\over 2}K_{ij}\, \dd z^i (f^j-i\d_n z^j) 
+{i\over 2} K_j \ \dd(f^j-i\d_n z^j)
\nonumber\\
&&\hskip3.mm 
+i\, \dd z^i \left( w_i-{1\over 4} K_i^{lm} \pb_l\pb_m\right)  
+{i\over 4} K_{ilm} \dd z^i \p^l\p^m 
+{i\over 2} K_{jk} \dd\p^j\p^k\Bigg\} 
+ h.c.
\ea
Comparing with (\ref{chiralbsf}) one sees that the combinations $(f^j-i\d_n z^j)$ and $(\fb_j+i\d_n\zb_j)$ are exactly those appearing in the boundary superfields. On the other hand, $\Sigma^{(2)}$ contains a $\d_n\zb_j$ without the corresponding $\fb_j$ which is the reason why we cannot rewrite the sum $\Sigma^{(2)}+\wh\Sigma$ in terms of boundary superfields and their variations. Indeed, as it stands, $\Sigma^{(2)}+\wh\Sigma$ is not supersymmetric. We now use the algebraic field equations for the auxiliary fields. They read
\be\label{feoms}
K^i_j f^j=w^i-\half K^i_{jk}\p^j\p^k
\quad , \quad
K_i^j \fb_j=w_i -\half K_i^{jk}\pb_j\pb_k \ ,
\ee
which allows us to rewrite $w_i-{1\over 4} K_i^{lm} \pb_l\pb_m=K_i^j\fb_j+{1\over 4} K_i^{lm} \pb_l\pb_m$. Then
\ba\label{SShateom}
\hskip-8.mm 
\big(\Sigma^{(2)}+\wh\Sigma\big)\Big\vert_{f-{\rm eom}}\hskip-3.mm
&=&{i\over 2}\Bigg\{ \dd z^i \Big[
K_i^j\,  (\fb_j+i\d_n\zb_j) + K_{ij}\,  (f^j-i\d_n z^j) 
+\half K_i^{lm}\,\pb_l\pb_m +\half K_{ilm}\, \p^l\p^m
\nonumber\\
&&
+\,K_{il}^m\, \p^l\s^n\pb_m \Big]
+\, K_i\, \dd (f^i-i\d_n z^i) +K_i^j\, \dd\p^i\s^n\pb_j + K_{ij}\, \dd\p^i\p^j\Bigg\} \ +h.c.
\ea
This can now be compactly rewritten in terms of the boundary superfields $\f^i,\ \bar\f_i$, cf (\ref{chiralbsf}), and their variations as
\be\label{SShatbsf}
\big(\Sigma^{(2)}+\wh\Sigma\big)\Big\vert_{f-{\rm eom}}
=\intt \left( 
-{i\over 2}\, \dd\f^i\, K_i\left(\f^j,\bar\f_k\right) 
+{i\over 2}\, \dd\bar\f_i\, K^i\left(\f^j,\bar\f_k\right) \right) \ ,
\ee
which now is manifestly supersymmetric.
It is perhaps useful to stress that this boundary term is a function of the boundary superfields only and {\it not} of their (normal) derivatives.  Of course, when expanding in components, $\d_n z$ is present in the boundary term, see (\ref{SShateom}). It is important that the appearance of $\d_n z$ is entirely determined by the expansion of the boundary superfield $\f$, cf (\ref{chiralbsf}). This results in a more rigid structure of the boundary term than is the case for a non-supersymmetric theory like the scalar example discussed in the previous subsection. 
Note that for a canonical K\"ahler potential and a single chiral superfield e.g. one simply has
$\big(\Sigma+\wh\Sigma\big)\Big\vert_{f-{\rm eom}}
=\intt \left( -{i\over 2}\, \dd\f\, \bar\f 
+ {i\over 2}\, \dd\bar\f\, \f\right) \ .
$

As discussed in the previous section, we have the freedom to add to  $S=S^{(2)}+\wh S$ a susy invariant boundary action $S_B$ that could depend a priori on the boundary superfields $\f^i$ and $\bar\f_i$, as well as their (super)derivatives. However, it is easy to convince oneself that any ($2+1$ Poincar\'e invariant) occurrence of such derivatives in $S_B$ will either lead to boundary terms involving second normal derivatives like $\d_n^2 z$ or involving $\d_\mh \d^\mh z$ (which is related by the bulk field equations to $\d_n^2 z$). Such higher derivative boundary terms will require corresponding ``higher-derivative" boundary conditions (involving e.g. $\d_n\f$, and hence $\d_n^2 z$ in components) which are inconsistent, as discussed above. Hence, we only allow to add a boundary action of the form (\ref{SBdef}), with the function $B$ only constrained by a reality condition:
\be\label{Breality}
S_B=\intdm \intt B(\f^l,\bar\f_k)
\quad ,\qquad 
\left[ B(z^l,\zb_k)\right]^* =  B(z^l,\zb_k) \ .
\ee
Variation of $S_B$ then yields additional boundary terms: $\dd S_B=\intdm \Sigma_B$ with
\be\label{SBvar}
\Sigma_B=\intt \left( B_i(\f^l,\bar\f_k) \, \dd \f^i+B^i(\f^l,\bar\f_k)\, \dd \bar\f_i\right)
\ee
Combining with (\ref{SShatbsf}) the complete boundary term reads
\be\label{allbounterms}
\big(\Sigma^{(2)}+\wh\Sigma\big)\Big\vert_{f-{\rm eom}}\hskip-1.mm+\,\Sigma_B
=\intt \hskip-1.mm\left[
\dd\f^i\left(B_i(\f^l,\bar\f_k)  -{i\over 2}\, K_i\left(\f^j,\bar\f_k\right) \right)
+\dd\bar\f_i \left( B^i(\f^l,\bar\f_k) +{i\over 2}\, K^i\left(\f^j,\bar\f_k\right) \right) \right]\ .
\ee

\subsection{Boundary conditions \label{bcs}}

Boundary conditions must relate the $\bar\f_j$ and the $\f^i$ in such a way that (\ref{allbounterms}) vanishes. Obviously, such boundary conditions will be  manifestly supersymmetric.

\subsubsection{Wess-Zumino model}

To begin with, let us look at the simplest case of a single chiral field with canonical K\"ahler potential and $B=0$ (Wess-Zumino model).
As noted above, in this case (\ref{allbounterms}) reduces to $\intt \left( -{i\over 2}\, \dd\f\, \bar\f 
+ {i\over 2}\, \dd\bar\f\, \f\right)$ which should vanish by the boundary condition. The latter must be of the form
\be\label{singleBC}
\bar\f= G(\f) \ .
\ee
Indeed, the only  condition not relating $\bar\f$ to $\f$ that could lead to a vanishing of the boundary terms
would be $\dd\f=\dd\bar\f=0$, implying that $\f$ is constant on the boundary and, hence, $z,\p$ and $\d_n z+if$ are all constant on the boundary. Since the auxiliary field $f$ has to be replaced by $\bar{w}'$ in the end, this would imply that $z$ and $\d_n z$ are fixed on the boundary, which is too strong and would eliminate all dynamics in the bulk. Thus any physically acceptable supersymmetric boundary condition must relate $\f$ and $\bar\f$. Such a condition can always be solved, at least locally, in the form (\ref{singleBC}). Furthermore, the function $G$ cannot be arbitrary. Taking the hermitian conjugate of (\ref{singleBC}) we see that the inverse function must equal the complex conjugate function:
\be\label{GGinverse}
G^{-1}=G^* \ .
\ee
We will call functions $G(z)$ that satisfy this relation {\it admissible} functions.\footnote{
Sometimes, one has to impose further conditions on $G$. For example, $G(z)={\gamma\over z}$ does satisfy (\ref{GGinverse}), but the condition $\zb=G(z)$ only makes sense for $\g>0$.
}
Now, (\ref{singleBC}) implies $\dd\bar\f= G'(\f) \dd\f$ and we see that the boundary term vanishes provided $(-G(\f)+G'(\f) \f)\dd\f=0$. We have just seen that $\dd\f$ cannot be zero, and we get a simple linear differential equation for the function $G$ in one (complex) variable:
\be\label{Gdiffsimple}
G(z) = z\, G'(z) \ ,
\ee
with admissible solution $G(z)=e^{i\dd} z$, so that the boundary condition  is
\be\label{bcsingle}
\bar\f= e^{i\dd} \f \ .
\ee

\subsubsection{$N$ superfields with arbitrary K\"ahler potential and $B=0$\label{NarbK}}

Let us now discuss the general supersymmetric sigma model but not yet adding the extra boundary term, i.e. keeping $B=0$.
In this case the boundary conditions must be such that (\ref{SShatbsf}) vanishes. This will require $N$ relations between the $\f^i$ and the $\bar\f_j$, which we could write as $g_{(i)}(\f^j,\bar\f_k)=0,\ i=1,\ldots N$. Possibly after taking appropriate linear combinations, the $g_{(i)}$ are real-valued functions.
Again, these conditions must be such that we can solve them (at least locally) to express all $\bar\f_j$ as functions of the $\f^k$, or else a subset of the $\f^i$ would satisfy some condition $f(\f^i)=0$ which would again lead to too strong boundary conditions. Hence, at least in principle, we can write the boundary conditions as
\be\label{severalBC}
\bar\f_i = G_i(\f^j)
\ee
Again, the complex conjugate functions must give the inverse relations. This constraint is best written in terms of the matrices of partial derivatives
\be\label{consistseveral}
G_{i,j} (G_{j,k})^*=\dd_i^k \quad , \quad G_{i,j}={\d G_i\over \d z^j} \ .
\ee
The boundary conditions (\ref{severalBC}) imply for the field variations
$\dd\bar\f_i=G_{i,j}\, \dd\f^j$ 
so that the boundary term  (\ref{SShatbsf}) vanishes if 
\be\label{sevGcond}
K_i\big(z^l,G_k(z^l)\big)- G_{m,i}(z^l)\ K^m\big(z^l,G_k(z^l)\big)=0 \ .
\ee
For given K\"ahler potential $K$, these are $N$ partial (in general non-linear) differential equations for the $N$ function $G_m$. A simple consistency  condition can be obtained by taking the derivatives with respect to $z^j$ and antisymmetrizing in $i$ and $j$. This yields the following integrability condition on the boundary conditions and K\"ahler metric\footnote{
A sufficient condition on the K\"ahler potential for (\ref{sevGcond}) and hence also (\ref{KGcons}) to be satisfied is
$K(z^l,\zb_k)=K(G_l^*(\zb), G_k(z))$,
stating that the K\"ahler potential is ``invariant" under $\zb_j \to G_j(z),\ z^i\to G^*_i(\zb)$. Indeed, taking ${\d\over \d z^i}$ of this relation and setting $\zb_k=G_k(z)$ yields (\ref{sevGcond}).
}
evaluated at $\zb_m=G_m(z^l)$:
\be\label{KGcons}
\Big(K^k_i\, G_{k,j} \Big)\Big\vert_{\zb_m=G_m(z^l)} = \Big(K^k_j\,  G_{k,i}\Big)\Big\vert_{\zb_m=G_m(z^l)} \ .
\ee
This integrability condition actually has a nice geometrical interpretation: $K_i^j$ is the K\"ahler metric on the scalar manifold which is the $N$-complex dimensional manifold with ``complex coordinates" $z^i$ et $\zb_i$. The K\"ahler 2-form is given by ${\cal K}=K_i^k \ {\rm d} z^i\wedge {\rm d} \zb_k$. The boundary conditions $\zb_i=G_i(z^j)$ constitute $N$ real conditions and hence select an $N$-real dimensional submanifold ${\cal L}$.  The pullback of the K\"ahler form to this submanifold then is 
\be\label{Lagrangiansub}
{\cal K}\Big\vert_{\cal L}=K_i^k \ {\rm d} z^i\wedge G_{k,j} {\rm d} z^j \Big\vert_{\zb_m=G_m(z^l)}={1\over 2} \left( K_i^k G_{k,j}-K_j^k G_{k,i}\right) \Big\vert_{\zb_m=G_m(z^l)} {\rm d} z^i\wedge {\rm d} z^j \ .
\ee
We see that the integrability condition (\ref{KGcons}) is equivalent to the vanishing of ${\cal K}\Big\vert_{\cal L}$ which is the statement that the submanifold ${\cal L}$ defined by the boundary conditions is a {\it Lagrangian submanifold}.

Let us note that the entire discussion of boundary conditions
is manifestly covariant with respect to holomorphic field redefinitions $\wt\Phi^i=F^i(\Phi^j)$ (and $\bar{\wt\Phi}_i=F_i(\bar\Phi_j)$ with $F_i\equiv \bar F^i\equiv (F^i)^*$). Indeed, the K\"ahler potential simply transforms as $\wt K(\wt z^l,\bar{\wt z}_k) = K(z^l,\zb_k)$ so that $K_i = M_i^{\ j}\, \wt K_j$ and 
$K^m= \overline M^m_{\ \, n}\, \wt K^n$ where
$M_i^{\ j}\equiv M(z)_i^{\ j}={\d F^j(z)\over \d z^i}$ and
$\overline M^m_{\ \,n}\equiv\overline M(\zb)^m_{\ \,n}= \overline{(M(z))_m^{\ \, n}}$.
The boundary condition $\zb_i=G_i(z^j)$ becomes $\bar{\wt z_i}=\wt G_i(\wt z^l)$ where $\wt G_i(\wt z^l)
=F_i\Big(G_j\big( (F^{-1})^k(\wt z^l)\big)\Big)\equiv \wt G_i(\wt z^l)$. It follows that $\wt G_{i,l}={\d\over \d \wt z^l} \wt G_i$ is given by $\wt G_{i,l} = (M^{-1})_l^{\ k}\, G_{j,k}\, \overline M^j_{\ i}$ or equivalently
$G_{j,k}=M_k^{\ l}\, \wt G_{i,l}\, (\overline M^{-1})^i_{\ j}$
so that indeed the l.h.s. of the boundary condition (\ref{sevGcond}) transforms as
\be
K_i - G_{m,i} K^m = M_i^{\ j} \Big( \wt K_j - \wt G_{n,j} \wt K^n\Big) \ .
\ee

Actually one can use this freedom to do holomorphic field redefinitions to go to ``coordinates" in which the boundary conditions are simply $\bar{\wt z}_i=\wt z^i$. Said differently, we can find new holomorphic coordinates such that the Lagrangian submanifold determined by the boundary conditions is simply characterized by vanishing imaginary parts of the $\wt z^i$. Of course,  this might be at the price of generating some complicated $\wt K_i$, $\wt K^m$, etc. To achieve this, it is enough to let
\be\label{trivbc}
\wt z^i \equiv F^i(z^l) = \l z^i + \bar\l G_i(z^l) \ , \quad \l\in {\bf C} \ ,
\ee
where $\l$ should be chosen such that this is invertible (except possibly at points where $F^i$ might have poles or cuts). Then, indeed, the condition $\zb_i=G_i(z^l)$ translates into $\bar{\wt z}_i=\wt z^i$. Note that this field redefinition (\ref{trivbc}) is not the only one that trivializes the boundary conditions: any $\wt{\wt z}^i=\wt F^i(\wt z^j)$, such that the $\wt{\wt z}^i$ are real whenever the $\wt z^i$ are real, does the job.

Let us finish this subsection by considering a large class of K\"ahler potentials of the form
\be\label{KcalK}
K(z^l,\zb_k)= k(z^l\zb_l)\ . 
\ee
This includes the canonical K\"ahler potential, as well as the standard K\"ahler potential on ${\bf CP}^N$.
Then $k'$ drops out of (\ref{sevGcond}) which reduces to $G_i=z^m G_{m,i}$. This is solved by
\be\label{GN}
G_j(z^l)=N_{jk}z^k \ ,\quad N^*=N^{-1} \ .
\ee
with a constant symmetric matrix $N$. In the simplest cases, $N$ is diagonal and then simply
$G_j(z^l)= e^{i\dd_j} z^j$ with the $\dd_j$ arbitrary constant phases.

\subsubsection{General boundary conditions from adding an invariant boundary action $S_B$\label{genbcsb}}

In this subsection, we will study the boundary conditions that can be obtained if we add the additional boundary action $S_B$. Now the boundary conditions $\bar\f_i=G_i(\f^j)$ must be such that the full boundary term (\ref{allbounterms}) vanishes.  Inserting $\delta\bar\f_i=G_{i,j}\,\delta\f^j$ into this boundary term we get
$B_i-{i\over 2} K_i+G_{j,i}(B^j+{i\over 2} K^j)=0$. It is understood that all quantities are evaluated with the boundary conditions imposed, i.e.  $K_i\equiv K_i(\f^l, G_k(\f^l))$, etc. Hence, these are $N$ partial differential equations in the $N$ variables $\f^i$. Equivalently, we can call these variables just $z^i$. Thus, after a slight rearrangement, our equations finally read
\be\label{sevGBcond}
{\d\over \d z^i} B\Big(z^l, G_k(z^l)\Big) = {i\over 2} \left[
K_i\Big(z^l, G_k(z^l)\Big)- G_{m,i}(z^l)\ K^m\Big(z^l, G_k(z^l)\Big)
\right] \ ,
\ee
or, equivalently
\be\label{sevGBcondequiv}
{\d\over \d z^i} \left[ 
B\Big(z^l, G_k(z^l)\Big) +{i\over 2}  K\Big(z^l, G_k(z^l)\Big) \right]
= i \,
K_i\Big(z^l, G_k(z^l)\Big) \ ,
\ee
Again, there is an integrability condition: if we take ${\d\over \d z^j}$ and antisymmetrize in $i$ and $j$, the left hand side obviously vanishes. For the right hand side, this implies the same condition (\ref{KGcons}) as we already derived above for vanishing $B$. Again, the submanifold determined by the boundary conditions must be a Lagrangian submanifold. 

If this integrability condition is satisfied, we can find a function $\wt B(z^l)$ such that its derivatives give the right hand side of (\ref{sevGBcond}). The non-trivial question is whether this function $\wt B(z^l)$ can be written as $B(z^l,G_k(z^l))$ with $B(z^l,\zb_k)$ real for any complex $z^l$. This is indeed the case and the proof is given in appendix B. Actually, in general, there are many ways to extend $B(z^l,G_k(z^l))$ defined on the Lagrangian submanifold to a real $B(z^l,\zb_k)$ on the whole K\"ahler manifold.
Thus given any set of boundary conditions that defines a Lagrangian submanifold of the K\"ahler manifold, we can find a family of appropriate real functions $B(z^i,\zb_i)$ and corresponding boundary actions $S_B$. 

Conversely, for given $B$, an admissible solution $\{ G_j(z^l)\}_{j=1,\ldots N}$ must now satisfy (\ref{consistseveral}). We will show again in appendix B that we can always find a solution of the differential equation (\ref{sevGBcond}) that satisfies (\ref{consistseveral}).
Actually, since the differential equations are non-linear, one typically has several ``branches" of solutions.\footnote{
This is somewhat similar to what happened for a single scalar field with $b=0$ where the non-linear condition $\d_n\vf\, \delta\vf=0$ has two branches of solutions: $\d_n\vf=0$ (Neumann) and $\vf=const$ (Dirichlet).
} 
We will see some explicit examples below.

Note again that our equations (\ref{sevGBcond}) and (\ref{sevGBcondequiv}) are perfectly covariant under arbitrary holomorphic field redefinitions. In particular, $\wt B(\wt z^l, \bar{\wt z}_k) = B(z^l, \zb_k)$ and one finds again that both sides of the equations transform with the same matrix $M$ defined above. As before, the field redefinition (\ref{trivbc}) allows us to always achieve $\overline{\wt \f}_i=\wt \f^i$.

Let us summarize: we have shown how the boundary conditions on the superfields are determined via (admissible) solutions of some differential equations that involve the K\"ahler potential and the function $B$ of the boundary action. We have shown that for any real $B$ one gets  corresponding supersymmetric boundary condition(s), and that for any given supersymmetric boundary condition that defines a Lagrangian submanifold, one can always find a (family of) corresponding real $B$.

\subsection{Examples of susy boundary conditions\label{examples}}

We will now give various more or less non-trivial examples of boundary conditions determining  appropriate boundary actions and, conversely, of boundary actions determining the boundary conditions. We have already discussed above  boundary conditions corresponding to a vanishing boundary action $S_B$. Most often, we will simply write the boundary conditions as $\zb_i=G_i(z^j)$ but one should keep in mind that this is a relation for the full boundary superfields: $\bar\f_i=G_i(\f^j)$.

\subsubsection{A single chiral superfield with canonical K\"ahler potential\label{single-canonicalK}}

We first look at the simplest case of a single chiral superfield with canonical K\"ahler potential $K(z,\zb)=\zb z$, i.e. the Wess-Zumino model to which we add the boundary action $S_B$.
The differential equation (\ref{sevGBcond}) then simply is
\be\label{Bdifcond}
{{\rm d}\over {\rm d} z} B\big(z, G(z)\big)= {i\over 2} \left( G(z)-z\, G'(z)\right) \ .
\ee

\noindent
{\bf\ref{single-canonicalK}.1}

\vskip2.mm
\noindent
Generically, we do not know how to solve the differential equation (\ref{Bdifcond}) for arbitrary (real) $B(z,\zb)$. However, if 
\be\label{BFzzb}
B(z,\zb)=F(|z|^2) \ ,
\ee
with a real function $F$, equation (\ref{Bdifcond}) becomes
$
{{\rm d}\over {\rm d} z} \left[ -i F(z\, G(z)) + {1\over 2} z G(z)\right] = G(z) \ .
$
Introducing $\wh G(z)=z\, G(z)$, this is immediately integrated as
$-i\int^{\wh G} F'(\xi) {{\rm d}\xi\over \xi} +\half \log\wh G=\log z$. Hence the boundary condition is
\be\label{fbHf}
\bar\f = \f\ \exp\left( 2 i H(\f \bar\f)\right)
\quad , \quad {\rm with}\ \ H(x)=\int^x F'(\xi){{\rm d}\xi\over \xi} \ .
\ee
This is like (\ref{bcsingle}) except that now the phase $\dd$ no longer is a constant but a function of the real superfield $\f\bar\f$.
Conversely, given a boundary condition like this with any real function $H(x)$, the appropriate boundary action is given by (\ref{BFzzb}) with
$F(x)=\int^x H'(\xi) \xi {\rm d}\xi$.
\vskip2.mm
\noindent
{\bf\ref{single-canonicalK}.2}

\vskip2.mm
\noindent
Suppose we have a linear boundary condition
\be\label{bclin}
a\f+\bar{a} \bar\f =\g \quad (\g\in {\bf R})
\quad\Rightarrow\quad G(z)={\g\over \bar a} - {a\over \bar a} z \ ,
\ee
Equation (\ref{Bdifcond}) is trivial to integrate and one gets $B(z,G(z))=i {\g\over 2\bar a} z + const$. Although this does not look real, we know that it actually is, provided we choose the constant appropriately. Indeed, from the condition $\zb=G(z)$ we have ${z\over\bar a}= {\g\over |a|^2}-{\zb\over a}$, so that we can rewrite
$B(z,\zb)\vert_{\zb=G(z)} = i{\g\over 4}\left({z\over \bar a} -{\zb\over a}\right)\vert_{\zb=G(z)}  + i {\g^2\over 4 |a|^2} 
+\b\left({z\over \bar a} +{\zb\over a}\right)\vert_{\zb=G(z)}-{\b\g\over |a|^2}+const$. Thus, up to the irrelevant constant, this defines a real function of $z$ and $\zb$. Hence we arrive at
\be\label{Blin}
B(z,\zb)= c z + \bar{c} \zb  \quad {\rm with} \quad {\rm Im}(\bar a c)={\g\over 4} \ .
\ee
Note that there is a whole family of functions $B$ since 
${\rm Re}(\bar a c)$ is not constrained. Of course, for $\g=0$, the boundary condition is $G(z)=-{a\over\bar a}z\equiv e^{i\dd} z$ and we can take $c=0$, consistently with (\ref{bcsingle}). 

\vskip2.mm
\noindent
{\bf\ref{single-canonicalK}.3}

\vskip2.mm
\noindent
Next, consider a non-linear (quadratic) boundary condition like $z^2-\zb^2+i=0$, which we choose to solve as $\zb=G(z)=\sqrt{z^2+i}$. Inserting this in the right hand side of eq.~(\ref{Bdifcond}) and integrating gives $B(z,G(z))=-\half \int^z {{\rm d}\z\over \sqrt{\z^2+i}}=-\half \log\Big( z+\sqrt{z^2+i}\Big)=-\half \log(z+G(z))$. Thus, $B(z,\zb)=-\half \log(z+\zb)$. Somewhat more generally, one finds that the boundary condition
\be\label{quadraticcond}
a(z+b)^2 + \bar{a} (\zb+\bar{b})^2=\g \ ,
\ee
\vskip-3.mm
\noindent
with real $\g$ corresponds to
\be\label{quadraticB}
B(z,\zb)={i\over 2}(b\zb-\bar{b} z) - {\g\over 2|a|} \log\left[ \sqrt{-ia} (z+b) +\sqrt{i\bar a} (\zb+\bar b)\right] \ ,
\ee
up to an irrelevant additive real constant. Note that in the limit $\g\to 0$, (\ref{quadraticcond}) reduces to a linear boundary condition and, indeed, (\ref{quadraticB}) reduces to (\ref{Blin}).
\vskip2.mm
\noindent
{\bf\ref{single-canonicalK}.4}

\vskip2.mm
\noindent
Finally, consider a boundary condition of the form 
\be\label{modulusbc}
|z+b|^2=\g \quad \Leftrightarrow \quad 
\zb +\bar{b} = {\g\over z+b} \qquad (\g>0) \ .
\ee
Then $G(z)= {\g\over z+b}-\bar{b}$. Inserting this into
the right hand side of eq.~(\ref{Bdifcond}) and integrating gives $B(z,G(z))={i\over 2} \left( {\g\, b\over z+b} - \bar{b} z + 2\g \log(z+b)\right)+const$. Using ${\g\over z+b}=\zb+\bar{b}$ we can rewrite this as
\be\label{modulusB}
B(z,\zb)={i\over 2} (b \zb-\bar b z)-{i\over 2}\g \log{\zb+\bar b\over z+b} \ + \b (\zb+\bar b)(z+b)\ ,  \quad \b,\g \in {\bf R} \ ,
\ee
which is manifestly real. Note again that there is a family of functions $B$ parametrised by $\b$.

\vskip2.mm
\noindent
{\bf\ref{single-canonicalK}.5}

\vskip2.mm
\noindent
Let us now take the function $B$ of the previous example and see whether
(\ref{modulusbc}) is the only corresponding boundary condition. To simplify things, take $b=\b=0$ so that 
\be\label{modulusB2}
B(z,\zb)=-{i\over 2}\g \log{\zb\over z} \ .
\ee
The differential equation (\ref{Bdifcond}) for $G(z)$ then becomes
$\g\left({G'\over G}-{1\over z}\right)+G-z G'=0$. If we let $G(z)=\g z\, f(z)$, this becomes $ f'\left( z^2-{1\over f}\right)=0$. Clearly, there are two branches of solutions. The first is $f={1\over z^2}$, i.e. $G(z)={\g\over z}$ as expected, while the second is $f'=0$, i.e. $G(z)=c\, z$ with $c=e^{i\delta}$:
\be\label{twosolutions}
{\rm solution\ 1\ : }\quad G(z)={\g\over z}
\quad , \qquad \hskip1.cm
{\rm solution\ 2\ : }\quad G(z)=e^{i\delta}\, z \ .
\ee
Clearly, if $\g<0$, only solution 2 is acceptable.

\subsubsection{A single chiral superfield with a non-canonical K\"ahler potential\label{single-noncanK}}

The differential equation (\ref{sevGBcondequiv}) now reduces to
\be\label{Bdifcond3}
{{\rm d}\over {\rm d} z} \Big[ B\big(z, G(z)\big)
+{i\over 2} K\big(z,G(z)\big) \Big]
= i\, K_z\big(z,G(z)\big)\ ,
\ee
where $K_z={\d\over \d z}K(z,\zb)$. 

\vskip2.mm
\noindent
{\bf\ref{single-noncanK}.1}

\vskip2.mm
\noindent
Consider a  K\"ahler potential $K$ and boundary function $B$ of the forms
\be\label{KBonefield1}
K(z,\zb)={\cal K}(|z|^2) \quad , \quad B(z,\zb)=F(|z|^2) \ .
\ee
The standard K\"ahler potential on ${\bf CP}^1$ e.g. is of this form. In analogy with the case of a canonical K\"ahler potential studied above (eqs.(\ref{BFzzb}) to (\ref{fbHf})) we find a boundary condition
\be\label{fbHKf}
\bar\f = \f\ \exp\left( 2 i H(\f \bar\f)\right)
\quad , \quad {\rm with}\ \ H(x)=\int^x {F'(\xi)\over {\cal K}'(\xi)}{{\rm d}\xi\over \xi} \ .
\ee
\vskip2.mm
\noindent
{\bf\ref{single-noncanK}.2}

\vskip2.mm
\noindent
Next, consider a  K\"ahler potential $K$ and boundary condition of the forms
\be
K(z,\zb)={\cal K}(|z|^2) \quad , \quad \zb={\g\over z}\equiv G(z) \ .
\ee
Equation (\ref{Bdifcond3}) then is easily integrated with the result
$B(z,G(z))=i\g {\cal K}'(\g)\log z + const$, so that
\be
B(z,\zb)= {i\over 2} \g {\cal K}'(\g) \log{z\over \zb} \ .
\ee
\vskip2.mm
\noindent
{\bf\ref{single-noncanK}.3}

\vskip2.mm
\noindent
Now, consider the simple linear boundary condition $a\f+\bar a \bar\f =\g$ for the standard K\"ahler potential  on ${\bf CP}^1$. Since the K\"ahler potential
is invariant under the redefiniton $z'=e^{i\dd/2}z,\ \zb'=e^{-i\dd/2}\zb$, we can make $a$ purely imaginary by such a redefinition, and then absorb $|a|$ by redefining $\g$. Thus 
\be
K(z,\zb)=\log(1+z\zb)  \quad , \quad G(z)=z+i\g \ .
\ee
Integrating (\ref{Bdifcond3}) yields $B(z,G(z))=-{i\over 2} {\g\over\sqrt{4+\g^2}}\log{ 2z+i\g+i\sqrt{4+\g^2}\over 2z+i\g-i\sqrt{4+\g^2}}$, which we rewrite, using $ z +i\g=\zb$ in the manifestly real form
\be
B(z,\zb)=-{i\over 2} {\g\over\sqrt{4+\g^2}}\log{  z +\zb +i\sqrt{4+\g^2}\over  z +  \zb-i\sqrt{4+\g^2}} \ .
\ee
Note that in the limit where $\g=0$, we correctly get a boundary condition $\zb=z$  and a vanishing boundary action, $B=0$, as discussed above (eqs (\ref{KcalK})  for ${\bf CP}^N$.

\vskip2.mm
\noindent
{\bf\ref{single-noncanK}.4}

\vskip2.mm
\noindent
One can discuss again the quadratic boundary condition (\ref{quadraticcond}) of \ref{single-canonicalK}.3 by taking advantage of our freedom to do arbitrary holomorphic field redefinitions. For simplicity, we let $a=i$, $b=0$ and $\g=1$, i.e. the boundary condition is $z^2-\zb^2+i=0$.
Setting $\wt \Phi= \Phi^2+{i\over 2}$ (and hence also for the boundary superfield $\wt\f=\f^2+{i\over 2}$) transforms the quadratic boundary condition into the ``trivial" linear one: $\overline{\wt\f} = \wt\f\equiv \wt G(\wt\f)$. 
The price one pays is that the new K\"ahler potential no longer is canonical but $\wt K(u,\bar u) = K(z,\zb)=z\zb=\sqrt{(u-{i\over 2})(\bar u+{i\over 2})}$ and $\wt K_u=\half\sqrt{\bar u+{i\over 2}\over u-{i\over 2}}$. The differential equation (\ref{Bdifcond3}) then yields
${{\rm d}\over {\rm d}u} \wt B\big(u,u\big)={i\over 2} \left[ \wt K_u\big(u,u\big)-\wt K^{\bar u}\big(u,u\big)\, \wt G'(u)\right]
= -{1\over 2} {1\over \sqrt{1+4u^2}}$,
with solution
$\wt B(u,\bar u)= -{1\over 2} \log\left[ \sqrt{u-{i\over 2}}+\sqrt{\ub+{i\over 2}}\ \right]
= -{1\over 2} \log\left[z+\zb \right]\equiv B(z,\zb)$,
which is exactly the $B$ found in 
{\ref{single-canonicalK}.3}, demonstrating explicitly for this example the covariance of our differential equation (\ref{Bdifcond3}) under holomorphic field redefinitions. 

\subsubsection{Several chiral superfields with arbitrary K\"ahler potential}

We will only look at one rather symmetric class of examples, generalizing (\ref{KBonefield1}) and (\ref{fbHKf}):
\be\label{KBsev}
K(z^l,\zb_k)={\cal K}(z^j\zb_j) \quad , \quad B(z^l,\zb_k)=F(z^j\zb_j) \ .
\ee
This includes, in particular, the standard K\"ahler potential on ${\bf CP}^N$. The differential equation (\ref{sevGBcond}) or (\ref{sevGBcondequiv}) then yields 
$\left( {F'(\xi)\over {\cal K}'(\xi)} +{i\over 2}\right) {\d \xi\over \d z^j} = i G_j$, where $\xi=z^j G_j(z^l)$.
The ansatz $G_j(z^l)= e^{2i H(\xi)} N_{jk} z^k$, with a symmetric and unitary constant matrix $N$, is a solution provided
$H(\xi)=\int^\xi  {F'(x)\over {\cal K}'(x)} {{\rm d}x\over x}$.
Thus, the boundary condition reads
\be\label{fbHNf}
\bar\f_j= \exp\left[2i H(\f^l \bar\f_l)\right]\ N_{jk}\, \f^k
\quad , \quad
H(\xi)=\int^\xi  {F'(x)\over {\cal K}'(x)} {{\rm d}x\over x}
 \ .
\ee
In particular, $\zb_j=N_{jk} z^k$ for $B\equiv F=0$ and we recover (\ref{GN}).

\subsection{Energy momentum conservation and boundary conditions\label{emcsusy}}

Just as for the scalar example discussed in the beginning of this section, we will end it by showing also in the supersymmetric case that the supersymmetric boundary conditions found above are exactly what is needed to ensure conservation of the total energy and of the two tangential components of the total momentum.

Again, translational invariance in all four space-time directions of the bulk Lagrangian ${\cal L}^{(2)}=\half \left[ K(\Phi^i, \bar\Phi_j) \right]_D+\left[ w(\Phi^i) \right]_F +\left[ \ov{w}(\bar\Phi_j) \right]_{\bar F}$ as given by eqs (\ref{Dtermexpl}) and (\ref{wcomp}), together with the bulk Euler-Lagrange field equations imply
\be\label{tmnsusy1}
\d_\m T^\m_{\ \, \n}=0 \quad , \quad T^\m_{\ \, \n} = {\d{\cal L}^{(2)}\over \d\, \d_\m\xi^a} \d_\n\xi^a - \dd^\m_\n\, {\cal L}^{(2)} \ ,
\ee
where $\xi^a$ stands generically for all fields. Explicitly we have
\be\label{tmnsusy2}
T^\m_{\ \, \n} = \left( K_i^j\, \d^\m z^i \d_\n\zb_j
-{i\over 2} K_i^j\, \p^i \s^\m \d_\n \pb_j
+{i\over 2} K^k_{ij}\, \p^j\s^\m \pb_k \d_\n z^i \right)  + h.c.
-\dd^\m_\n\, {\cal L}^{(2)} \ .
\ee
In analogy with our discussion of the scalar field case, we expect the conserved total energy and conserved total tangential momentum to be given by
\be\label{susyPconserved}
P^\nh=\int_{x^n\le 0} {\rm d}^3 x\, T^{0\nh} 
- g^{0\nh} \int_{x^n=0} {\rm d}^2 x\, \left(\wh{\cal L} +\intt B(\f^i,\bar\f_j) \right) \ ,
\ee
where $\wh{\cal L}$ is the integrand of the boundary action $\wh S$ we had to add to the standard bulk action $S^{(2)}$, cf eq. (\ref{Shat}), namely
\be\label{Lhat}
\wh{\cal L}={i\over 2} \left[ K_i (f^i-i\d_n z^i) +{1\over 2} K_{ij}\p^i\p^j +2 w\right] + h.c. \ .
\ee
Indeed, for $\nh= 0$ we expect the total energy to receive contributions from the ``boundary potentials" which are $-\wh{\cal L}$ and $ -\intt B$.

Let us now verify that the $P^\nh$ defined in (\ref{susyPconserved}) are indeed conserved. Using $\d_\m T^\m_{\ \, \n}=0$ we have
\be\label{susycons1}
{{\rm d}\over {\rm d} t} P^\nh = \int_{x^n=0} {\rm d}^2 x\, \left[ -T^{n\nh} - g^{0\nh} \d_0 \left(\wh{\cal L} +\intt B(\f^i,\bar\f_j) \right)\right] \ .
\ee
To show that this vanishes, consider the equality of the right-hand sides of eqs~(\ref{SShateom}) and (\ref{SShatbsf}) and replace $\dd\f^i$ by $\d_\nh\f^i$, $\dd z^i$ by $\d_\nh z^i$, etc. After some rearrangements and using the auxiliary field equations of motion (\ref{feoms}), the resulting equality can be written as
\be\label{susycons2}
\intt \left( 
-{i\over 2}\, \d_\nh\f^i\, K_i\left(\f^j,\bar\f_k\right) 
+{i\over 2}\, \d_\nh\bar\f_i\, K^i\left(\f^j,\bar\f_k\right) \right) 
=\d_\nh \wh{\cal L} + T^n_{\ \,\nh}\Big\vert_{x^n=0} \ .
\ee
Using the boundary conditions $\bar\f_i=G_i(\f^l)$ and the partial differential equations (\ref{sevGBcond}) which determine them, the left-hand side of (\ref{susycons2}) is seen to be $-\d_\nh \intt B\Big(\f^i, G_j(\f^l)\Big)$, so that on the boundary
\be\label{susycons3}
T^n_{\ \,\nh} = -\d_\nh \left(\wh{\cal L} +\intt B(\f^i,G_j(\f^l)) \right) \qquad {\rm on} \ \d\cM \ .
\ee
It follows that, provided the boundary conditions are satisfied, the integrand of (\ref{susycons1}) vanishes for $\nh=0$, while for $\nh\ne 0$ it is a total derivative in the boundary plane and the integral again vanishes. Hence $P^\nh$ as defined by (\ref{susyPconserved}) is indeed conserved.

\section{Susy couplings to new boundary fields\label{newboundfields}}
\setcounter{equation}{0}

So far, we have given a rather complete description of the non-linear sigma model in the presence of a boundary. This involved the bulk superfields $\Phi^i$ and corresponding boundary superfields $\f^i$, as well as their hermitian conjugate fields. We introduced possible susy boundary terms of these same boundary superfields $\f^i$. We did not, however, introduce any new degrees of freedom on the boundary.

It is the purpose of this section to show that it is quite straightforward to also introduce new boundary superfields $\vf^A$ on the boundary and to couple them to the $\f^i$, in a supersymmetric way. We will first quickly recall the $2+1$ dimensional sigma model action of the $\vf^A$ alone, and then study the effect of possible couplings to the $\f^i$.

\subsection{Susy sigma-model in 2+1 dimensions\label{ssm3d}}

As discussed in section \ref{BSBS},
a general boundary superfield has the expansion (cf. (\ref{genbsf}))
\be\label{genbsf2}
\vf(x^\mh,\t)=\z(x^\mh) +\rd \t\c(x^\mh) - \t\t g(x^\mh) \ .
\ee
Such a boundary superfield can be complex or real, $\bar\vf=\vf$ (in components: $\bar\z=\z$, $\bar g = g$ and $\bar\c = \c\s^n$). In the complex case, the real and imaginary parts transform irreducibly under the $2+1$ dimensional super Poincar\'e group and we can consider them separately as real. We will assume henceforth that all $\vf^A$ are real boundary superfields.
The $2+1$ dimensional supersymmetry generators $\wt Q$ were given in (\ref{susygenerators}): $\wt Q_\a = -i {\d\over \d\t^\a} + 2 \,\wh\g^\mh_{\a\b} \t^\b {\d\over \d x^\mh}$ with $\wh\g^\mh_{\a\b}= \wh\g^\mh_{\b\a}=- \s^\mh_{\a\gd} \e^{\gd\dot\dd}\s^n_{\b\dot\dd}$. They anticommute with the components of the superderivative
\be\label{3dsuperderiv}
\wt D_\a={\d\over \d\t^\a} - 2i \,\wh\g^\mh_{\a\b} \t^\b {\d\over \d x^\mh} \ .
\ee
Acting with this superderivative on a boundary superfield gives another boundary superfield. In particular, it can be used to construct the ``kinetic superfield" as\footnote{
The $2+1$ dimensional Dirac matrices $\G^\mh$  are such that $\c\G^\mh\p\equiv\c^\a\G_\a^{\mh\b}\p_\b=\c^\a\wh\g^\mh_{\a\g}\p^\g\equiv\c\wh\g^\mh\p$. They satisfy $\G^\mh \G^\nh=\s^\mh \sb^\nh$ and, hence, by (\ref{clifford}), the $2+1$ dimensional Clifford algebra.
}
\be\label{3dkinetic}
-{1\over 4} D^\a D_\a \vf
= -g -\rd \t(i\G^\mh \d_\mh\c) - \t\t\, \d^\mh\d_\mh \z\quad ,
\quad (\G^\mh)_\a^{\ \b} =\wh\g^\mh_{\a\g}\e^{\g\b} \ ,
\ee
The $2+1$ dimensional susy sigma-model action for these ``new" boundary superfields $\vf^A$ is given as
\be\label{3daction}
S_3=S_3^{\rm kin}+S_3^\w \ ,
\ee
\vskip-3.mm
\noindent
with
\ba\label{3dkin}
S_3^{\rm kin}&=&
-{1\over 4} \int{\rm d}^3 x\intt H_A(\vf^B) D^\a D_\a \vf^A
\nonumber\\ 
&=& \int{\rm d}^3 x\  h_{AB}  \Big(
\d^\mh \z^A\d_\mh \z^B + i \c^A\G^\mh D_\mh \c^B
+g^A g^B + \Gamma^A_{CD} g^B \c^C\c^D \Big) \ ,
\\ \label{3dw}
S_3^\w&=&2\int{\rm d}^3 x\intt \w(\vf)
=\int{\rm d}^3 x \left(-2\w_A g^A - \w_{AB} \c^A\c^B\right)\ .
\ea
Here $\w$ is a real boundary superpotential and $h_{AB}$ the (real) metric  on the boundary field scalar manifold, given by
$h_{AB}={1\over 2} (H_{A,B}+H_{B,A})$,
where we used the standard notation $H_{A,B}=\partial_B H_A$. 
Furthermore, $\Gamma^A_{CD}$ are the Christoffel symbols (not 
to be confused with the $\G^\mh$ matrices) associated with this metric
and they are simply given by  
$\Gamma_{CD}^A={1\over 2} h^{AB}H_{B,CD}$. They
also appear in the covariant derivative
$D_\mh \c^B=\d_\mh\c^B+\G^B_{CD}\d_\mh \z^C\c^D$. 
The kinetic terms provide boundary propagators for the boundary fields $\z^A$ and $\c^A$, while the
$g^A$ are non-propagating auxiliary fields. If one integrates out these auxiliary fields one gets the on-shell action as can be found e.g. in \cite{NdW}.

\subsection{Coupling the new boundary fields to the bulk fields\label{couplnew}}

As extensively discussed and used above, the bulk (anti) chiral superfields $\Phi^i$ (and $\bar\Phi_i$) give rise to boundary superfields $\f^i$ (and $\bar\f_i$).
We have written down a boundary term $S_B$ for these boundary superfields $\f^i$ and $\bar\f_i$ which has exactly the same structure as the boundary superpotential term $S_3^{\w}$ defined in (\ref{3dw}): Indeed, if we define $\w_{(B)}\Big( \f^i+ \bar\f_i,\, i(\f^i- \bar\f_i)\Big)\equiv \half B(\f^i, \bar\f_i) $ then $\w_{(B)}$ is a real boundary superpotential having the real and imaginary parts of $\f^i$ as its arguments. On the other hand, we did not add any boundary kinetic term like $S_3^{\rm kin}$ for the $\f^i$ or $\bar\f_i$ since we did not want the bulk fields to have an independent propagator on the boundary. We also argued that  such boundary kinetic terms for the $\f^i$ or $\bar\f_i$ would lead to boundary terms involving second normal derivatives and are inconsistent variational principle.

In the presence of the new boundary fields $\vf^A$ it is natural to combine $B(\f^i,\bar\f_j)$ and $\omega(\vf^A)$ into a single function $B(\f^i,\bar\f_j,\vf^A)$ so that
\be\label{SBwithvarphi}
S_B=\intdm\intt\, B(\f^i,\bar\f_j, \vf^A)\ .
\ee
The full action to consider then is
\be\label{fullaction}
S+\wh S + S_B + S_3^{\rm kin} \ ,
\ee
with $S+\wh S$ the minimal susy invariant  action determined in section 3, and $S_3^{\rm kin}$ given by (\ref{3dkin}).

We can now repeat the analysis of section \ref{bcandeom} and ask which boundary conditions will ensure stationarity of the action when the field equations are satisfied. First consider the new fields $\vf^A$. They are purely $2+1$ dimensional fields on $\d\cM$, and since $\d\cM$ has no boundary, vanishing of the $\dd\vf^A$-terms will simply lead to their $2+1$ dimensional field equations. Of course, the latter will also involve $\f^i$ and $\bar\f_j$. On the other hand,  requiring the variation of the action with respect to the $\Phi^i$ and $\bar\Phi_j$ to vanish, now leads to exactly the same bulk field equations as before, but also to boundary conditions that now can depend on the new boundary fields. Indeed, the vanishing of the boundary terms now relates the function $B(z^l,\zb_k,\z^B)$ to boundary conditions of the form
\be\label{bcvarphi}
\bar\f_j = G_j(\f^i,\vf^A)
\ee
via the differential equation
\be\label{sevGBcondvarphi}
{\d\over \d z^i} B\Big(z^l, G_k(z^l,\z^B),\z^B\Big) 
= {i\over 2} \left[
K_i\Big(z^l, G_k(z^l,\z^B)\Big)- G_{m,i}(z^l,\z^B)\ K^m\Big(z^l, G_k(z^l,\z^B)\Big)
\right] \ .
\ee
This equation is no more difficult to analyze in the presence of the new boundary superfields than in their absence. It is clear that the $\z^B$ appear in (\ref{sevGBcondvarphi}) and in the boundary condition $\zb_j=G_j(z^i,\z^A)$ only as additional real {\it parameters}. Thus they play no different role from any of the real parameters appearing in our examples like e.g. $\g$ in (\ref{quadraticcond}) and (\ref{quadraticB}) or in (\ref{modulusbc}) and (\ref{modulusB}).

For example, for a single chiral bulk superfield with a canonical K\"ahler potential and a single ``new" real boundary superfield $\vf$, the following choice of boundary action
\be\label{ffbvf1}
S_B=\intdm\intt\,\left[\,  -{i\over 2} \, \vf \log {\bar\f\over \f }\, \right]
\ee
leads to either of the two boundary conditions 
\be\label{ffbvf2}
\bar\f \, \f=\, \vf \quad {\rm or} \quad \bar\f=e^{i\dd}\, \f \quad\quad {\rm on}\  \ \ \d\cM\ .
\ee

To show the conservation of the total energy and tangential components of the total momentum proceeds similarly as in the previous section. Of course, the new fields $\vf^A$ now also have their own three-dimensional energy-momentum tensor $T^\nh_{(3)\mh}$ which needs to be included. More precisely, if we let $S_3^{\rm kin}=\intdm {\cal L}_{(3)}$ then $T^\nh_{(3)\mh}$ is computed from ${\cal L}_{(3)}$ only.
The fact that ${\cal L}_{(3)}$ does not depend explicitly on $x^\mh$, together with the  Euler-Lagrange field equations for $\vf^A$ imply
\be\label{T3mn}
\d_\mh T^\mh_{(3)\nh}=\intt \d_\nh \vf^A\, {\d \over \d \vf^A} B(\f^i,\bar\f_j,\vf^C) \ .
\ee
We continue to call $T^\m_{\ \,\n}$ the bulk energy-momentum tensor for the $\Phi^i$ and $\ov\Phi_j$ as defined in (\ref{tmnsusy2}). It still satisfies (\ref{susycons2}), but the boundary conditions now also involve the $\vf^A$ so that  (\ref{susycons3}) is modified to
\be\label{susycons3bis}
T^n_{\ \,\nh} = -\d_\nh \left(\wh{\cal L} +\intt B(\f^i,G_j(\f^l,\vf^C),\vf^C)\right)  + \intt \d^\nh \vf^A {\d\over \d \vf^A} B(\f^i,G_j(\f^l,\vf^C),\vf^C)
\quad {\rm on} \ \d\cM \ .
\ee
We then define
\be\label{susyconsnew}
P^\nh=\int_{x^n\le 0} {\rm d}^3 x\, T^{0\nh} 
+\int_{x^n=0} {\rm d}^2 x\, \left[ T^{0\nh}_{(3)}
- g^{0\nh} \left(\wh{\cal L} +\intt B(\f^i,\bar\f_j,\vf^A) \right)\right] \ .
\ee
Using $\d_\m T^{\m\nh}=0$, eqs~(\ref{T3mn}), (\ref{susycons3bis}) and $\int_{x^n=0}{\rm d}^2 x\, \d_{\hat t} T^{\hat t \nh}_{(3)} =0$ (where $\hat t$ labels the two tangential directions), as well as the boundary conditions, we get
\be\label{susyconsnew2}
{{\rm d}\over {\rm d} t} P^\nh = \int_{x^n=0} {\rm d}^2 x\, \left[ \d^\nh \left(\wh{\cal L} +\intt B\right)  - g^{0\nh}\, \d_0 \left(\wh{\cal L} +\intt B \right)\right] \ ,
\ee
where the terms involving $\d_\nh \vf^A \,{\d B\over \d \vf^A}  $ have cancelled. As before, we see that the $P^\nh$ are conserved.


\section{Permeable walls\label{permwalls}}
\setcounter{equation}{0}

We now want to apply our formalism to study junctions between two domains, say ${\cal D}_1=\{ x\in {\bf R}^4 \vert\, x^n\le 0\}$ and
${\cal D}_2=\{ x\in{\bf R}^4\vert\,x^n\ge 0\}$ that meet on a common boundary or wall which we will call ${\cal W}$, see Fig. \ref{permwalls-bis}.  Such a situation might be thought of as a generalization of the low-energy effective theory of the domain wall example mentioned in the introduction and studied further in appendix C. In each domain, we consider a collection of chiral superfields with some non-linear sigma-model action, and a priori unrelated K\"ahler metrics and superpotentials. We want to study under which conditions this combined system is still supersymmetric and, in particular, which are the supersymmetric matching conditions one can impose across the wall. One trivial possibility is to have just two separate sigma-models, each with its own boundary conditions, not involving the other fields, and the boundary action $S_B$ just being the sum  of the two independent boundary actions. Such boundary conditions might be called ``purely reflective" and the wall ``impermeable". In this section, we will study the more interesting case with ``(partly) transmissive" boundary conditions, where the two sets of fields interact through a permeable wall. This is achieved by taking a general boundary action $S_B$ that couples both types of boundary fields.

The present situation can be mapped, via a folding procedure, to the one discussed in the previous sections. More precisely, on can map all fields defined in ${\cal D}_2$ to some mirror fields defined in ${\cal D}_1$ and then study the system of the original fields and the mirror fields in the single domain ${\cal D}_1$ with boundary ${\cal W}$ as we did in the previous sections. Alternatively, one can just do a direct analysis of the coupling of the two sigma-models. While both lead to the same result, of course, we will only present the direct analysis which at present is a bit more straightforward.

We will end this section with several explicit and rather non-trivial examples, matching e.g. sigma models with different K\"ahler potentials on different scalar manifolds and with different superpotentials. Note that the whole discussion of the present section can also be immediately generalized to several domains ${\cal D}_p$ separated by {\it parallel} walls ${\cal W}_{p,p+1}$. 

\subsection{Analysis of the matching conditions\label{directanalysis}}

\begin{figure}[h]
\centering
\includegraphics[width=0.4\textwidth]{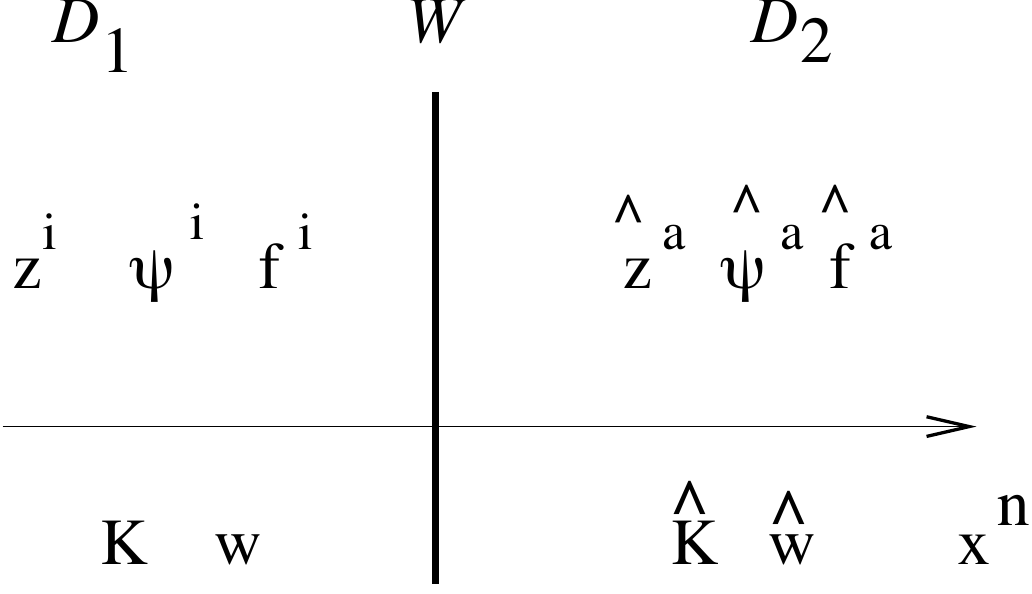}\\
\caption[]{Two different sigma-models living on adjacent space-time  domains ${\cal D}_{1}$ and
${\cal D}_{2}$ separated by a common boundary 
${\cal W}$.} 
\label{permwalls-bis}
\end{figure}

We continue to call $\Phi^i$ and $z^i, \p^i, f^i$ the (super) fields living 
on ${\cal D}_1$, i.e. ``on the left'', while we denote 
$\widehat\Phi^a$ and $\hz^a, \hp^a, \hf^a$ those living
on ${\cal D}_2$, i.e. ``on the right'', see Fig \ref{permwalls-bis}. 
Similarly, we write $K$ and $w$, resp. $\widehat K$ and $\widehat w$ for 
the K\"ahler potentials and superpotentials.

\noindent
To begin with, the total bulk action  then is
\ba\label{S12}
S&=&\int_{{\cal D}_1} {\rm d}^4x\, 
\left(\half \left[ K(\Phi^i, \bar\Phi_j) \right]_D
+\left[ w(\Phi^i) \right]_F 
+ \left[\bar{w}(\bar\Phi_i) \right]_{\bar{F}} \right)
\nonumber\\
&+&\int_{{\cal D}_2}   {\rm d}^4x\,
\left(\half \left[ \widehat K(\widehat\Phi^a, \bar{\widehat\Phi}_b) \right]_D
+\left[\widehat w(\widehat\Phi^a) \right]_F 
+ \left[\bar{\widehat w}(\bar{\widehat\Phi}_a) \right]_{\bar{F}} \right)\ .
\ea
Just as in section \ref{btfsv}, the susy variation of each of the two terms will produce a boundary term, but due to the opposite orientation of the 
boundary of ${\cal D}_2$ the second term comes with an opposite 
sign. Restricting ourselves as before to the supersymmetries satisfying $\eb=\e\s^n$, these boundary terms are canceled by the susy variation of an appropriate boundary action $\wh S$. As in section \ref{Shatsect}, one finds
\ba\label{ShatD1D2}
\wh S&=&- \int_{{\cal W}} {\rm d}^3 x\, \Im \left\{  
K_j \big(f^j-i\d_n z^j\big) +\half K_{jk}\p^j\p^k 
+2 w \right\} 
\nonumber\\
&&+\int_{{\cal W}} {\rm d}^3 x\, \Im \left\{  
\wh K_b \big(\wh f^b-i\d_n \wh z^b\big) +\half \wh K_{ab}\wh \p^a\wh \p^b 
+2 \wh w \right\} \ ,
\ea
and then
$\dd (S+\wh S) \Big\vert_{\eb=\e\s^n} =0$.
Again, one can add a boundary term $S_B$ that is susy invariant by itself and, hence, is written as a boundary superspace integral of an arbitrary real function of all boundary superfields: 
\be\label{SBwall}
S_B=\int_{{\cal W}} {\rm d}^3 x \intt B(\f^i,\bar\f_j, \wh\f^a, \overline{\wh\f}_b) \ ,
\ee 
The function $B$ can also depend on additional real superfields $\varphi^A$ living only on the boundary, as discussed in section \ref{newboundfields}. As we have seen there, for the discussion of the boundary conditions, these additional fields play the same role as real parameters appearing in the function $B$ and we will not indicate their possible presence explicitly. It is rather straightforward to generalize the discussion of supersymmetric boundary conditions to the present situation, resulting in a set of partial differential equations similar to (\ref{sevGBcond}). Indeed, the boundary conditions can now be expressed in terms of the boundary superfields as\footnote{The functions $G_i$ and $\wh G_a$ can also depend on the additional $\vf^A$ but, again, we do not write this out explicitly.}
\be\label{wallbc}
\bar\f_i=G_i(\f^j,\wh\f^b)
\quad , \quad
\overline{\wh\f}_a=\wh G_a(\f^j,\wh\f^b) \ .
\ee
Vanishing of the boundary terms when varying the action $S+\wh S+S_B$ (and using the auxiliary field equations of motion) then leads to
\ba\label{sevGBcondwall}
{\d\over \d z^i} B\Big(z^l, G_k, \wh z^c, \wh G_d\Big) &=& {i\over 2} \left[
K_i\Big(z^l, G_k\Big)- G_{m,i}\ K^m\Big(z^l, G_k\Big)
+ \wh G_{b,i}\ \wh K^b\Big(\wh z^c, \wh G_d \Big)
\right] 
\nonumber \\
{\d\over \d \wh z^a} B\Big(z^l, G_k, \wh z^c, \wh G_d\Big) &=& {i\over 2} \left[
-\wh K_a\Big(\wh z^c, \wh G_d\Big)+ \wh G_{e,a}\ \wh K^e\Big(\wh z^c, \wh G_d\Big)
- G_{m,a}\  K^m\Big(z^l, G_k \Big)
\right] 
\ ,
\ea
where $G_{m,i}\equiv G_{m,i}(z^l,\wh z^d)\equiv {\d\over \d z^i}G_m(z^l,\wh z^d),\ \wh G_{b,a}\equiv \wh G_{b,a}(z^l,\wh z^d)\equiv {\d\over \d \wh z^a}\wh G_b(z^l,\wh z^d)$, etc.

These partial differential equations again have integrability conditions. Taking ${\d\over \d z^j}$ of the first equation and antisymmetrizing in $i$ and $j$ gives back the condition (\ref{KGcons}). Similarly, taking ${\d\over \d z^b}$ of the second equation and antisymmetrizing in $a$ and $b$ gives an analogous condition on the $\wh G$ and $\wh K$: 
\be\label{intcond3}
\wh K^e_b\, \wh G_{e,a} = \wh K^e_a\, \wh G_{e,b} \ ,
\ee
where it is understood that the arguments are as in (\ref{sevGBcondwall})
More interestingly, taking ${\d\over \d z^a}$ of the first equation and subtracting ${\d\over \d z^i}$ of the second yields
\be\label{KKhatmatching}
\wh K^b_a  \wh G_{b,i} = - K^j_i G_{j,a} \ ,
\ee
again with all arguments as in (\ref{sevGBcondwall}). This equation provides a matching condition between the K\"ahler metrics on both sides of the wall.  Of course, if the boundary conditions do not mix the $\f^i$ and $\wh\f^a$, we have $\wh G_{b,i}=G_{j,a}=0$, and 
the matching condition (\ref{KKhatmatching}) is trivially satisfied. This corresponds to ``purely reflective" boundary conditions and an ``impermeable" wall. In the general case, however, where $G_i$ and $\wh G_a$ depend on both, $\f^j$ and $\wh \f^b$, the condition (\ref{KKhatmatching}) is non-trivial.
Note, that we do not get any matching condition on the superpotential. Thus, we can have completely different superpotentials on both sides of the wall.

Again, we can show that the boundary conditions ensure conservation of the appropriately defined total energy and tangential components of total momentum. They now get contributions from both domains:
\be\label{wallemt}
P^\nh= \int_{x^n\le 0} {\rm d}^3 x\, T^{0\nh}_{(1)} 
+ \int_{x^n\ge 0} {\rm d}^3 x\, T^{0\nh}_{(2)} 
-g^{0\nh} \int_{x^n=0} {\rm d}^2 x\, 
\left( \wh{\cal L} + \intt B \right) \ ,
\ee
where now $\wh{\cal L}$ is the integrand of (\ref{ShatD1D2}) and $T^{0\nh}_{(1)}$, resp. $T^{0\nh}_{(2)}$ are the obvious generalizations of (\ref{tmnsusy2}). Following exactly the same steps as in section \ref{emcsusy}, one then shows that
\be\label{walltmncons}
{{\rm d}\over {\rm d}t} P^\nh =0\ .
\ee

Instead of solving the boundary conditions to express the $\bar\f_i$ and $\overline{\wh\f}_a$ in terms of the $\f^j$ and $\wh\f^b$ as in (\ref{wallbc}), we can choose to solve them to express the  $\bar\f_i$ and ${\wh\f}^a$ in terms of the $\f^j$ and $\overline{\wh\f}_b$ as
\be\label{wallbc2}
\bar\f_i=\cG_i(\f^j,\ov{\wh\f}_b)
\quad , \quad
\wh\f^a=\wh \cG^a(\f^j,\ov{\wh\f}_b) \ .
\ee
This would be the natural choice when applying the folding procedure since the latter acts as parity (combined with a rotation) and hence exchanges dotted and undotted spinors. Thus by supersymmetry, the folding must map $\wh \f^a$ to some $\ov{\wt \f}_a$
Then, eqs (\ref{sevGBcondwall}) would be replaced by
\ba\label{sevGBcondwall2}
{\d\over \d z^i} B\Big(z^l, \cG_k, \wh \cG^c, \ov{\wh z}_d\Big) &=& {i\over 2} \left[
K_i\Big(z^l, \cG_k\Big)- \cG_{m,i}\ K^m\Big(z^l, \cG_k\Big)
- \wh \cG^b_{\ ,i}\ \wh K_b\Big(\wh \cG^c, \ov{\wh z}_d \Big)
\right] 
\nonumber \\
{\d\over \d \ov{\wh z}_a} B\Big(z^l,\cG_k,\wh \cG^c,\ov{\wh z}_d\Big) &=& {i\over 2} \left[
\wh K^a\Big(\wh \cG^c, \ov{\wh z}_d\Big)
-\wh\cG^{e,a} \wh K_e\Big(\wh \cG^c, \ov{\wh z}_d\Big)
- \cG_m^{\ \, ,a}\  K^m\Big( z^l, \cG_k\Big)
\right] 
\ ,
\ea
where now $\cG_{m,i}\equiv \cG_{m,i}(z^l,\ov{\wh z}_d)\equiv {\d\over \d z^i}\cG_m(z^l,\ov{\wh z}_d),\ \wh \cG^b_{\ ,i}\equiv \wh \cG^b_{\ ,i}(z^l,\ov{\wh z}_d)\equiv {\d\over \d z^i}\wh \cG^b(z^l,\ov{\wh z}_d)$, etc. In terms of the $\cG_i$ and $\wh\cG^a$ the integrability condition (\ref{KKhatmatching}) now is replaced by
\be\label{intcond4}
\wh K^a_b\, \wh \cG^b_{\ ,i} = K_i^j\, \cG_j^{\ ,a} \ .
\ee
The differential equations (\ref{sevGBcondwall2}) can be rewritten in an equivalent but more convenient form as
\ba\label{sevGBcondwall2bis}
{\d\over \d z^i} \Big[ B\Big(z^l, \cG_k, \wh \cG^c, \ov{\wh z}_d\Big) 
+ {i\over 2} K\Big(z^l, \cG_k\Big) 
+ {i\over 2} \wh K\Big(\wh \cG^c, \ov{\wh z}_d \Big) \Big]
&=& i\ K_i\Big(z^l, \cG_k\Big)
\nonumber \\
{\d\over \d \ov{\wh z}_a} 
\Big[ B\Big(z^l, \cG_k, \wh \cG^c, \ov{\wh z}_d\Big) 
+ {i\over 2} K\Big(z^l, \cG_k\Big) 
+ {i\over 2} \wh K\Big(\wh \cG^c, \ov{\wh z}_d \Big) \Big]
&=& i\ \wh K^a\Big(\wh \cG^c, \ov{\wh z}_d\Big) \ .
\ea
Note again that the functions $\cG_i$ and $\wh \cG_a$ cannot be arbitrary but must be such that eqs (\ref{wallbc2}) are consistent with their hermitian conjugate equations. This implies the generalization of eq.~(\ref{consistseveral}), namely
\be\label{consistfold}
N\, N^* = {\bf 1} \quad , \quad 
N= \pmatrix{
\cG_{i,j}& \cG_i^{\ ,b} \cr
\wh\cG^a_{\ ,j} & \wh\cG^{a,b}\cr } \   .  
\ee

In subsection \ref{NarbK} we remarked that one could always trivialize the boundary conditions, i.e. find a holomorphic field redefinition such that the boundary conditions become $\zb_i=z^i$. It is still true that, if one did the folding procedure, one could find such a holomorphic field redefinition. However, in general it would mix the $\f^i$ and the mirror images of the $\ov{\wh\f}_a$, i.e. fields living on different domains (before the folding). Physically this makes no sense and, in the present situation, the ``non-diagonal" $\cG_i^{\ ,b}$ and $\wh\cG^a_{\ ,j}$ carry physical information, namely they control the permeability  of the wall.

\subsection{Examples\label{Examples}}
%
It is not difficult to construct explicit examples of such supersymmetric junctions between different non-linear sigma models in domains ${\cal D}_1$ and ${\cal D}_2$. Here, we will first discuss the case of linear boundary (matching) conditions with canonical K\"ahler potentials on both sides (where $B=0$), and then the more interesting case  with different K\"ahler potentials on each side of the wall (where $B\ne 0$). We will find it convenient to use the formulation
(\ref{wallbc2})-(\ref{sevGBcondwall2bis}) of the boundary conditions and matching conditions for the K\"ahler metric. Note, once again, that if $\wh\cG^b_{\ ,i}=\cG_j^{\ ,a}=0$
the boundary conditions are ``purely reflective" and the analysis reduces to the one of section \ref{bcandeom} for two separate models. We will not discuss these cases further, but only note that even in this case one can have some minimal interaction between the two sigma-models by coupling the fields of each one to some common new boundary superfields $\vf^A$ via a boundary action $\int_{\cal W} {\rm d}^3 x \intt \left[ B_1(\f^i,\bar\f_j,\vf^A) + B_2(\wh\f^a, \ov{\wh\f}_b, \vf^A)\right]$, with $\vf^A$ having also a 3-dimensional kinetic term $S_3^{\rm kin}$. 

%
\subsubsection{Canonical K\"ahler potentials and same number of superfields in both domains}

Although this is a very simple example, it can still be very interesting since there is no matching condition on the superpotentials. Hence one can have completely different superpotentials in both domains. 

In subsection \ref{single-canonicalK} we have studied many linear and non-linear boundary conditions. Here, we will restrict ourselves to linear boundary conditions:
\be\label{walllincond}
\cG_i(z^j,\ov{\wh z}_b)=\cG_{i,j}\, z^j + \cG_i^{\ ,b}\, \ov{\wh z}_b
\quad , \quad 
\wh\cG^a(z^j,\ov{\wh z}_b)=\wh\cG^a_{\ ,j}\, z^j + \wh\cG^{a,b}\,\ov{\wh z}_b \ , 
\ee
with constant $\cG_{i,j}, \cG_i^{\ ,b}, \wh\cG^a_{\ ,j}, \wh\cG^{a,b}$. Then eqs. (\ref{sevGBcondwall2}) are satisfied with $B=0$, provided
\be\label{symmetryreq}
\cG_{i,j}=\cG_{j,i} \quad , \quad \wh\cG^{a,b}=\wh\cG^{b,a} \quad , \quad
\cG_i^{\ ,b}=\wh\cG^b_{\ ,i} 
\ee
which amounts to requiring that the matrix $N$ defined in (\ref{consistfold}) is symmetric (and, hence, unitary). Of course, eqs.~(\ref{symmetryreq}) also ensure that the integrability conditions like (\ref{intcond4}) are satisfied.

For a {\it single} chiral superfield with canonical K\"ahler potential in each domain, $N$ is a unitary, symmetric $2\times 2$ matrix $\pmatrix{a&b\cr b&d\cr}$. After appropriate phase rotations of $\f$ and of $\wh\f$, one can assume $a,d>0$. It is then easy to see that one can parametrize $N$ in terms of a single real parameter $\eta$ as
\be\label{Neta}
N={1\over 1+\eta^2} \pmatrix{ 1-\eta^2 & 2i\eta\cr 2i\eta& 1-\eta^2\cr} \qquad {\rm with} \quad  -1\le\eta\le 1 \ .
\ee
In terms of this parameter $\eta$, the present linear boundary conditions can be rewritten simply as
\be\label{etabc}
\wh\f-\ov{\wh\f}=i\eta ( \f+\ov\f)
\quad , \quad
\ov\f-\f=i\eta (\wh\f+\ov{\wh\f}) \ .
\ee
In particular, for the scalars this reads $\Im \wh z=\eta \Re z$ and $\Re \wh z=-{1\over \eta}\Im z$. For $\eta\ne\pm 1$ this implies a change in the complex structure between the two scalar manifolds.

\subsubsection{A single superfield in each domain with different K\"ahler potentials}

Now, we want to consider examples where the two theories in ${\cal D}_1$ and ${\cal D}_2$ have different K\"ahler potentials (as well as different superpotentials). To simplify the discussion, we will first restrict ourselves to a single chiral superfield in each domain. Also, to simplify the notation, we will write $\xi$ instead of $\wh z$ for the scalar field in ${\cal D}_2$. Furthermore, we will only consider K\"ahler potentials of the form
\be\label{KandKhat}
K(z,\zb)=\cK(z\zb) \quad , \quad \wh K(\xi,\xb)=\wh \cK(\xi\xb) \ .
\ee
For example, one can have the standard K\"ahler potential of ${\bf CP}^1$ in one domain and a canonical K\"ahler potential  in the other.

Proceeding in analogy  with the examples studied in eqs.~(\ref{BFzzb}) to (\ref{fbHf})) and eqs.~(\ref{KBonefield1}) to (\ref{fbHKf}) we make the ansatz 
\be\label{ffhatgghat}
\cG(z,\xb)= z \exp\left[ i\a\,\wh g\,\wh \cK'(\wh g) + 2i H(g)\right]
\quad , \quad 
\wh\cG(z,\xb)= \xb \exp\left[ i\a\, g\,\cK'(g)+2i \wh H(\wh g)\right] \ ,
\ee
where $\a$ is some real parameter and $g\equiv g(z,\xb)$ and $\wh g\equiv\wh g(z,\xb)$ are defined as
\be\label{gghat}
g= z\, \cG(z,\xb) \equiv z\,\zb
\quad , \quad
\wh g= \xb \wh\cG(z,\xb) \equiv \xi\,\xb \ ,
\ee
as well as $B=B(g,\wh g)$.
The differential equations (\ref{sevGBcondwall2bis}) relating  $\cG$, $\wh\cG$ and $B$ then read
\ba\label{sevGBcondwall2bissingle}
z{\d\over \d z} \Big[ B(g,\wh g) 
+ {i\over 2} \cK(g) + {i\over 2} \wh \cK(\wh g) \Big]
&=& i\ g\, \cK'(g)
\nonumber \\
\xb{\d\over \d \xb} 
\Big[ B(g,\wh g) 
+ {i\over 2} \cK(g) + {i\over 2} \wh \cK(\wh g) \Big]
&=& i\ \wh g\, \wh\cK'(\wh g) \ .
\ea
Using (\ref{gghat}) one easily finds
\be\label{fderfhatder}
z{\d\over \d z}\cK(g) =g\,\cK'(g) \left[2 + i\a\, z{\d\over \d z}\Big( \wh g\, \wh\cK'(\wh g)\Big)+ 2i H'(g)\, z{\d\over \d z} g\right]\ ,
\ee
as well as a similar relation for $\xb{\d\over \d \xb}\wh\cK(\wh g)$.
It follows that (\ref{sevGBcondwall2bissingle}) is solved by
\be\label{Bwallsingle}
B(g,\wh g)={\a\over 2} \, g\,\cK'(g)\; \wh g\, \wh\cK'(\wh g) +F(g) + \wh F (\wh g) \ ,
\ee
with
\be\label{FHKrelwall}
F(g)=\int^g \cK'(\l) H'(\l)\, \l {\rm d}\l
\quad\Leftrightarrow\quad
H(g)=\int^g {F'(\l)\over \cK'(\l)} {{\rm d}\l\over\l} \ ,
\ee
and an analogous relation between $\wh F(\wh g)$, $\wh\cK'(\wh g)$ and $\wh H(\wh g)$. Clearly, the parameter $\a$ controls the permeability of the wall and in the limit $\a\to 0$ the above equations reduce to those describing two non interacting theories on ${\cal D}_1$ and ${\cal D}_2$, cf.~eqs.~(\ref{KBonefield1}) to (\ref{fbHKf}).

\subsubsection{Several superfields in each domain with different K\"ahler potentials}

Just as eqs.~(\ref{KBonefield1}) to (\ref{fbHKf}) for a single superfield could be generalized to eqs.~(\ref{KBsev}) and (\ref{fbHNf}) for several superfields, we can generalize the previous example to the case of $N$ superfields  in each domain  with different K\"ahler potentials
generalizing (\ref{KandKhat}) (again, we will call $\xi^a$ the scalars in ${\cal D}_2$):
\be\label{KKhatsev}
K(z^l,\zb_k)=\cK(z^j\zb_j) \quad , \quad
\wh K(\xi^c,\xb_d)=\wh\cK(\xi^b\xb_b) \ .
\ee
We now make the ansatz
\be\label{ffhatgghat2}
\cG_i(z^k,\xb_b)= C_{ij}\, z^j \exp\left[ i\a\,\wh g\,\wh \cK'(\wh g) + 2i H(g)\right]
\quad , \quad 
\wh\cG^a(z^k,\xb_b)= D^{ab}\,\xb_b \exp\left[ i\a\, g\,\cK'(g)+2i \wh H(\wh g)\right] \ ,
\ee
where $\a$ is again a real parameter and now
\be\label{gghat2}
g= z^j\, \cG_j(z^k,\xb_b) \equiv z^j\,\zb_j
\quad , \quad
\wh g= \xb_a\, \wh\cG^a(z^k,\xb_b) \equiv \xi^a\,\xb_a \ .
\ee
Then eqs.~(\ref{sevGBcondwall2bissingle}) and (\ref{fderfhatder}) remain unchanged, except for the substitutions $z{\d\over \d z} \to z^i{\d\over \d z^i}$ and $\xb{\d\over \d \xb} \to \xb_a{\d\over \d \xb_a}\,$. Hence $B(g,\wh g)$ is still given by (\ref{Bwallsingle}) with $F$ determined by (\ref{FHKrelwall}) in terms of $\cK'$ and $H$, as well as an analogous relation for $\wh F$.

%
%
\section{Conclusions}
 
We have presented a quite comprehensive study of rigid unextended supersymmetry on $3+1$ dimensional space-times with boundaries. The boundaries preserve a $2+1$ dimensional super-Poincar\'e algebra which contains two out of the original four supersymmetries. 

We have identified the relevant boundary superspace and the minimal boundary term one has to add to the standard bulk action of the supersymmetric non-linear sigma model to make it off-shell invariant under these two supersymmetries, even in the presence of the boundary. No boundary conditions are imposed at this stage. Further, non-minimal supersymmetric boundary actions $S_B$ are boundary superspace integrals of real functions $B$ of the boundary superfields. Boundary conditions then arise from the variational principle and are determined as solutions of a certain set of non-linear (partial) differential equations involving the function $B$ and the first derivatives of the K\"ahler potential. These supersymmetric boundary conditions can be complicated and non-linear themselves. We proved that for any choice of supersymmetric boundary conditions one can construct an appropriate boundary action $S_B$, and we gave many explicit examples.
We have shown how to include and couple additional new superfields that only live on the boundary. Finally, we have generalized these results to the coupling of two different sigma models, with different K\"ahler potentials and superpotentials, living in adjacent domains, and worked out the matching conditions, again providing several explicit examples. In all cases we have shown that the boundary conditions ensure conservation of the appropriately modified total energy and  tangential components of the total momentum.

An obvious question that we did not address is that of spontaneous susy breaking by the boundary conditions: 
2+1 Poincar\'e invariant vacua would be configurations with the scalar field expectation values $z_0^i\equiv\, \langle z^i\rangle$ depending at most on $x^n$ and vanishing fermion fields. To preserve the supersymmetry, they must obey the  bulk conditions $w^j(\zb_0)=i\partial_n z_0^j$ and the susy boundary conditions determined by the boundary potential $B$. If all solutions of the bulk conditions  are incompatible with the susy boundary conditions we would have spontaneous susy breaking by the boundary conditions. It is easy to imagine a bulk superpotential for a single chiral superfield such that e.g. $|z_0(x^n=0)|=2\sqrt{\g}$ for some $\g>0$. This then is incompatible with the boundary condition $z \zb=\g$, cf. (\ref{modulusbc}) to (\ref{twosolutions}). However, the theory is defined by the triple $(K,B,w)$ and as seen in (\ref{twosolutions}), the same function $B$ given in (\ref{modulusB2}) also allows for the other boundary condition $\zb=z$ which is compatible with the bulk solution. We postpone to future the study and construction of models that do spontaneously break susy by incompatibility of the bulk and boundary conditions.

Furthermore, it would be most interesting to extend the analysis of the present paper to include gauge fields, extended supersymmetry, and ultimately present an analogous general analysis for the case of supergravity. 


\vskip 0.4cm
\centerline{\large \bf Acknowledgements}
\vskip .3cm
\noindent
The project of the present paper started a long time ago. I am deeply indebted to Costas Bachas and Jean-Pierre Derendinger for a very fruitful collaboration in the early stage, as well as numerous discussions in the sequel.
This work was partially supported by the EU under contracts MRTN-CT-2004-005104 and MRTN-CT-2004-512194,  by the french ANR grant ANR(CNRS-USAR) no.05-BLAN-0079-01, as well as the swiss national science foundation.

\vskip 1.0cm

\newpage

\noindent
{\Large \bf Appendix A: \
Spinor identities \\ \label{Conventions}}
\renewcommand{\theequation}{A.\arabic{equation}}
\setcounter{equation}{0}

\noindent
Our conventions for undotted and dotted two-component spinors $\p_\a$ and $\pb_\ad$ as well as for the matrices $\s^\m$ and $\sb^\m$ where given in section 2.1. Here, we will give a few identities that are useful in the main text. 

If one recalls that the spinors are anticommuting and that 
$(\p_\a \c_\b)^+=\bar\c_\b \pb_\a$ it is easy to show the following identities
\ba
\p\chi=\chi\p \quad , \quad \pb\cb &=& \cb\pb \quad , \quad
(\p\chi)^\dag = \pb\cb \nonumber\\
\chi\s^\m\pb = -\pb\sb^\m\chi \quad , \quad
\chi \s^\m \sb^\n \p &=& \p \s^\n \sb^\m \chi \quad , \quad
\cb\sb^\m\s^\n\pb = \pb\sb^\n\s^\m\cb\nonumber\\
(\chi\s^\m \pb)^\dag = \p \s^\m \cb \quad &,& \quad
(\chi\s^\m \sb^\n\p)^\dag=\pb\sb^\n\s^\m\cb 
\ . 
\label{spinorrelations}
\ea
The $\s$ and $\sb$ matrices satisfy a ``Clifford algebra''
\be
\sb^\m \s^\n + \sb^\n \s^\m
= \s^\m \sb^\n + \s^\n \sb^\m = 2 g^{\m\n} {\bf 1} \ ,
\label{clifford}
\ee
as well as $(\s^j)^T=(\s^j)^*=i\s_2\ \s^j\ i\s_2$.

When evaluating superspace integrals the following identities are useful
\ba\label{thetaidentities}
\t^\a\t^\b=-\half \e^{\a\b}\,\t\t \quad , \quad
\t_\a\t_\b=\half \e_{\a\b}\,\t\t \quad &,& \quad
\tb^\ad\tb^\bd=\half \e^{\ad\bd}\, \tb\tb \quad , \quad
\tb_\ad\tb_\bd=-\half\e_{\ad\bd} \,\tb\tb \ ,\nonumber\\
\t\p\ \t\chi=-\half \t\t\ \p\chi \quad &,& \quad
\tb\pb\ \tb\cb =-\half \tb\tb\ \pb\cb \ ,\nonumber\\
\t\s^\m\tb\ \t\s^\n\tb =\half\t\t\, \tb\tb\ g^{\m\n} \quad &,& \quad
\t\s^\m\pb\ \t\s^\n\cb =\half \t\t\ \pb\sb^\m\s^\n\cb \ .
\ea
A general superfield has the expansion
\ba
\hskip-1.cm
S(x,\t,\tb)&=& C+i\t\chi-i\tb\cb +\t\s^\m\tb v_\m
+ {i\over 2} \t\t (M+iN) -{i\over 2} \tb\tb (M-iN)
\nonumber\\
&&+ i\, \t\t\, \tb\left( \bar\l +{i\over 2}\sb^\m\d_\m\chi\right)
- i\, \tb\tb\, \t\left( \l -{i\over 2}\s^\m\d_\m\cb\right) 
+\half \t\t\tb\tb \left( D-\half \d_\m\d^\m C\right) 
\label{gensfield}
\ea
where all component fields depend only on $x^\m$ : $C\equiv C(x)$ etc.
The superderivatives
\be\label{superderiv}
D_\a={\d\over\d\t^\a} + i \s^\m_{\a\bd}\tb^\bd {\d\over\d x^\m}
\quad , \quad
\ov{D}_\ad={\d\over\d\tb^\ad}+i\t^\b\s^\m_{\b\ad}\,{\d\over\d x^\m}
\ee
anticommute with the supersymmetry generators $Q_\b$ and $\ov{Q}_\bd$ given in (\ref{deltagensfield}) and allow us to impose supersymmetric constraints on general superfields. Chiral superfields $\Phi$ obey $\ov{D}_\ad\Phi=0$ and thus have the expansion given in eqs.~(\ref{chiral}) and (\ref{delta}). The susy variation for a general superfield $S(x,\t,\tb)$ is \cite{abgif}
\be
\dd S \equiv (i\e Q+i\eb \overline Q) S
={\d\over \d x^\m} \left( (i\t\s^\m\eb-i\e\s^\m\tb) S\right)
+{\d\over \d \t^\a} \left( -\e^\a S\right)
+{\d\over \d \tb^\ad} \left( -\eb^\ad S\right) \ ,
\label{Fsusyvar}
\ee
while for a chiral superfield $\Phi(y,\t)$ it is given in eq. (\ref{phisusyvar}).

\vskip8.mm
\newpage

\noindent
{\Large \bf Appendix B: \ Constructing real $B$  and admissible solutions}
\renewcommand{\theequation}{B.\arabic{equation}}
\setcounter{equation}{0}

\vskip3.mm
\noindent
In this appendix, we will show that, for given boundary conditions, the function $B(z^i, G_i(z^j))$ obtained by integrating the differential equations (\ref{sevGBcond}), can always be extended to a real function $B(z^i,\zb_i)$ on the entire scalar manifold. We will also show that for any given real $B$ and solution $G_i$ of the differential equations, one can always choose the constants of integration such that $G_i$ is an admissible solution.

\vskip3.mm
\noindent

We first look at the simplest case of a single chiral superfield with canonical K\"ahler potential $K(z,\zb)=\zb z$, i.e. the Wess-Zumino model to which we add the boundary action $S_B$.
Recall that the differential equation then simply is (\ref{Bdifcond}) \be\label{Bdifcondbis}
{{\rm d}\over {\rm d} z} B\big(z, G(z)\big)= {i\over 2} \left( G(z)-z\, G'(z)\right) \ ,
\ee
For a given boundary condition, i.e. given function $G(z)$, one can integrate the right hand side of this equation 
as
\be\label{diffsol}
B\big(z, G(z)\big)=\wt B(z) \quad , \quad {\rm with} \quad
\wt B(z)={i\over 2}\int^z (G(x')-x' G'(x')) {\rm d} x' \ .
\ee
The non-trivial question is whether one can find a function $B(z,\zb)$ which is real for all $z\in {\bf C}$ and which is such that $B(z,G(z)) = \wt B(z)$. For the examples studied in the main text, it was more or less easy to see how to use the condition $\zb=G(z)$ to rewrite the function $\wt B(z)$ as an obviously real function of $z$ and $\zb$.  We will now show that one can always obtain such a real function $B(z,\zb)$. The proof also provides a general explicit construction, although for specific examples this might not be the simplest one. We have already seen that the function $B(z,\zb)$ is not unique at all: indeed, if the boundary condition $\zb=G(z)$ is the solution of $g(z,G(z))=0$ with real $g$,  we can always add $g(z,\zb)$ (or any non-singular function of $g(z,\zb)$) to $B(z,\zb)$.

Proof: Let us first show that the function $\wt B(z)$ defined by the integral (\ref{diffsol}) is real whenever $z$ satisfies $\zb=G(z)$. More precisely, $\wt B$ (and also $B$) is only defined up to an arbitrary additive constant which has to be fixed appropriately.  The condition $\zb=G(z)$ defines a curve (possibly with several disconnected components) in the complex plane which we parametrize\footnote{
For example, writing $z=x+i y$, the boundary condition $z^2+ \zb^2=2\g$ yields two hyperbolas $y=\pm\sqrt{x^2-\g}$ and, locally on each, we can use $x$ as parameter to write $z=x\pm i \sqrt{x^2-\g}$ and $\zb= x\mp i \sqrt{x^2-\g}$.
} 
by a real parameter $x$ as $z=h(x)$ and $\zb=\bar h(x)$.
Note that, by definition, we  have $\bar h(x) = G(h(x))$, and hence also $G'(h(x))=\bar h'(x) / h'(x)$.
We need to show that $\wt B(h(x))$ is real, up to a constant, i.e. that ${{\rm d}\over {\rm d} x} \wt B(h(x))$ is real. Using (\ref{diffsol}) we get
\ba\label{realityproof1}
{{\rm d}\over {\rm d} x} \wt B(h(x)) 
&=& h'(x) {{\rm d}\over {\rm d} z} \wt B(z) \Big\vert_{z=h(x)}
= {i\over 2}\, h'(x) \Big( G(z)-z G'(z) \Big)\Big\vert_{z=h(x)}
\nonumber\\
&=&{i\over 2}\, h'(x) \Bigg( \bar h(x) -h(x) {\bar h'(x)\over h'(x)} \Bigg)
=\Im \Big( h(x) \bar h'(x)  \Big) \ ,
\ea
which is real. Adjusting the additive constant, we then have a real $\wt B(h(x))$. This defines a real function on the curve determined by the boundary condition. To define a real $B(z,\zb)$ for all complex $z$, we note that (if $x$ is a good parameter along the curve) one can invert the function $h$ (and hence $\bar h$) {\it on the curve}, so that $x=h^{-1}(z)$ and $x=\bar h^{-1}(\zb)$. One can then analytically continue this function $h^{-1}$ (and hence $\bar h^{-1}$) away from the curve. Then, of course, $h^{-1}(z)$ and $\bar h^{-1}(\zb)$ no longer are real and equal, but complex conjugate. Define
\be\label{BBtildedef}
B(z,\zb)=\wt B\left( h\left( {1\over 2}\left( h^{-1}(z)+\bar h^{-1}(\zb) \right) \right) \right) \ .
\ee
For any complex $z$ this is real since it equals $\wt B(h(\hat x))$ with a real $\hat x$, and for $\zb=G(z)$ it reduces to $\wt B(h(x))=\wt B(z)$. This completes the proof. 

Note that the analytically continued function $h^{-1}(z)$ we constructed during the proof actually provides a field redefinition $\wt{\Phi}=h^{-1}(\Phi)$ such that, in terms of the $\wt\Phi$, the boundary condition is simply the linear condition $\overline{\wt\Phi} = \wt\Phi$ (i.e. $\overline{\wt z}=\wt z$ for $\wt z=h^{-1}(z)$). 

We have shown how to construct a real $B(z,\zb)$ for any admissible $G(z)$. Conversely, we will now show that for a given real $B(z,\zb)$ and solution $G(z)$ of the differential equation, one can always choose the integration constants such that $G(z)$ is an admissible solution, i.e. a function $G$ such that $G^{-1}=G^*$.  Suppose that $g(z)$ is a solution of (\ref{Bdifcondbis}). Let $f=g^{-1}$. Let us show that $\bar f$ is a solution of the same differential equation. To do so, we start from (\ref{Bdifcondbis}), divide it by $g'(z)$, use $g$ as independent variable and write $z=f(g)$. Then
\be\label{hdiffcond}
{{\rm d}\over {\rm d} g} B(f(g),g)={i\over 2} \left( g {{\rm d} f(g)\over {\rm d}g} - f(g) \right)
\ee
Take the complex conjugate and note that the reality of $B(z,\zb)$ implies $[B(u,v)]^*=B(\bar v, \bar u)$. Then
\be\label{hdiffcond2}
{{\rm d}\over {\rm d} \bar g} B(\bar g, \bar f(\bar g))=-{i\over 2} \left( \bar g {{\rm d} \bar f(\bar g)\over {\rm d}\bar g} - \bar f(\bar g) \right)
= + {i\over 2} \left( \bar f(\bar g) -\bar g\ \bar f'(\bar g) \right)
\ .
\ee
If we now write $z$ instead of $\bar g$ for the independent variable, we see that $\bar f$ satisfies exactly the same equation (\ref{Bdifcond}) as $g$. Upon choosing the constant of integration appropriately, we can then take\footnote{
Of course, the fact that two functions satisfy the same {\it non-linear} first-order differential equation does not  imply that they are identical, up to a choice of integration constant. Indeed, we have seen the example of $G(z)=c z$ and $G(z)={\g\over z}$ being both solutions of the same equation. However, these two solutions are not of the form $g$ and $\ov{g^{-1}}$.
}
$\bar f=g$, i.e. $g$ is an admissible solution.

For a single chiral field with an arbitrary K\"ahler potential the differential equation (\ref{Bdifcondbis}) is replaced by (\ref{Bdifcond3}), or equivalently
\be\label{Bdifcond2bis}
{{\rm d}\over {\rm d} z} B\big(z, G(z)\big)= {i\over 2} \Big[ K_z\big(z,G(z)\big)-K^{\zb}\big(z,G(z)\big)\, G'(z)\Big] \ ,
\ee
It is completely straightforward to adapt the above proof to this case.
The only difference is that (\ref{realityproof1}) now is modified to read (note that $K_z(z,\zb)$ is the complex conjugate of $K^{\zb}(z,\zb)$)
\ba\label{realityproof2}
\hskip-1.cm {{\rm d}\over {\rm d} x} \wt B(h(x)) 
&=& h'(x) {{\rm d}\over {\rm d} z} \wt B(z) \Big\vert_{z=h(x)}
= {i\over 2}\, h'(x) \Big[ K_z\Big(z,G(z)\Big)-K^{\zb}\Big(z,G(z)\Big) \,G'(z) \Big]\Big\vert_{z=h(x)}
\nonumber\\
&&\hskip-2.cm ={i\over 2}\, h'(x) \Bigg[ K_z\Big(h(x),\bar h(x)\Big) -K^{\zb}\Big(h(x),\bar h(x)\Big) {\bar h'(x)\over h'(x)} \Bigg]
=\Im \Big[ K^{\zb}\Big(h(x),\bar h(x)\Big)\ \bar h'(x) \Big] \ .
\ea
The rest of the proof is as above.

Finally, the generalization to an arbitrary number of chiral fields with arbitrary K\"ahler potential is no more difficult, except for the notations that become involved. Hence, we will not spell it out here.

\vskip5.mm
%
\noindent
{\Large \bf Appendix C: \ Integrating out a heavy superfield around a domain wall\\ \label{domainwall}}
\renewcommand{\theequation}{C.\arabic{equation}}
\setcounter{equation}{0}

\noindent
In this appendix we will give some details concerning the example mentioned in the introduction -- as a motivation -- of integrating out the fluctuations of a heavy superfield around a domain wall solution.
Recall that  we consider a simple  model consisting of a heavy chiral superfield $\Phi_{\rm H}$ and a light one $\Phi$, with canonical K\"ahler potential and a superpotential given by
\be\label{FiFitildesup}
W(\Phi_{\rm H},\Phi)=\g (\Phi_{\rm H}^3/3-M^2\Phi_{\rm H}) + \Phi_{\rm H}\, w_2(\Phi) + w_1(\Phi)\quad , \quad w_1'(0)=w_2(0)=w_2'(0)=0 \ .
\ee 
The supersymmetric vacua for the heavy field $\Phi_{\rm H}$, given by $\d W/\d \Phi_{\rm H} =0$, then are $z_{\rm H}=\pm M$, where $z_{\rm H}$ denotes the scalar component of $\Phi_{\rm H}$. There also is the domain wall solution \be\label{dwsol}
z_{\rm H}(x^n)= M \, \tanh \left( \g M x^n\right) \ ,
\ee 
that interpolates between the two vacua and preserves two out of the initial four supersymmetries.  This domain wall has a width $(\g M)^{-1}$. Its tension (energy per unit area) is $\sim \g M^3$. The excitations around this solution can be seen to have a mass $\sim \g M$. We will be interested in the limit where $M$ is fixed and $\g \to\infty$. This corresponds to the thin wall limit and infinitely heavy fluctuations, but also to strong coupling. Nevertheless, we expect to be able to integrate out these fluctuations of the heavy superfield and to obtain an effective action for the light superfield, preserving the same two supersymmetries. This effective action should have an effective superpotential $W(-M,\Phi)= w_1(\Phi)-M w_2(\Phi) + const$ for $x^n<0$ and $W(M,\Phi)= w_1(\Phi)+M w_2(\Phi) + const$ for $x^n>0$.
We do not know how to show this rigorously, since the usual holomorphicity arguments require four preserved supersymmetries. However, far away from the domain wall at $x^n=0$, the physics should look as if the heavy field were in one of its vacua. 

The effective superpotential for the field $\Phi$ obtained by integrating out the field $\Phi_{\rm H}$ around the {\it vacuum} $z_{\rm H}=\pm M$ can be obtained exactly in the limit we are interested in. Holomorphicity together with a non-anomalous (there are no gauge fields here) $U(1)_R$-charge assignment $r_M=r_{\Phi_{\rm H}}$, $\ r_{w_2}=2-r_{\Phi_{\rm H}}$, $\ r_\g=2-3r_{\Phi_{\rm H}}$ (and $r_{w_1}=2$) restrict the effective superpotential $w_{\rm eff}^\pm(\Phi)$ to be of the form 
\be\label{weff}
w_{\rm eff}^\pm(\Phi)= M w_2\ f_\pm \left( { w_2(\Phi)\over \g M^2}\right) + w_1(\Phi) \ .
\ee 
The important point is that the symmetry considerations constrain  $w_{\rm eff}^\pm$ to depend on $\g$ only through the combination $\g M^2$. This will allow us to obtain a result at strong coupling from a weak coupling analysis. Indeed, to determine the function $f_\pm(\l)$ one can consider perturbation theory of $\Phi_{\rm H}$ around the given vacuum where (dropping an additive constant)
$W(\pm M +\Phi_{\rm H}',\Phi)=\pm \g M (\Phi_{\rm H}')^2 +{\g\over 3} (\Phi_{\rm H}')^3  +\Phi_{\rm H}' w_2 \pm M w_2 + w_1$. In the limit $\g\to 0$ with $\g M$ fixed (this is $\l\equiv{w_2\over \g M^2}\to 0$), the superfield $\Phi_{\rm H}'$ becomes a free superfield and doing the Gaussian integration yields
$w_{\rm eff}^\pm=\mp {w_2^2\over 4 \g M} \pm M w_2 +w_1+{\cal O}(\g)$, so that we idenify 
\be\label{ffct}
f_\pm (\l) =\pm \left( 1-{\l\over 4} + {\cal O}(\l^2)\right) \ .
\ee 
We are interested in a different limit, however, with $M$ fixed and $\g\to\infty$. Nevertheless, since this also yields $\l\equiv{w_2\over \g M^2}\to 0$, one can still use the small $\l$ expansion of $f_\pm$ and one gets
\be\label{wefffinal}
w_{\rm eff}^\pm(\Phi)=w_1(\Phi) \pm M w_2(\Phi) \mp {w_2(\Phi)^2\over 4 \g M} +{\cal O}({1\over \g^2}) \ .
\ee
The leading (finite) terms coincide indeed with the naive result one gets from simply substituting the classical value $\pm M$ for $\Phi_{\rm H}$ in $W(\Phi_{\rm H},\Phi)$.



\begin{thebibliography}{99}

\bibitem{early}
H.~Luckock,
{\it Boundary terms for globally supersymmetric actions},
Int.\ J.\ Theor.\ Phys.\  {\bf 36}, 501-508 (1997).
  
\bibitem{sigma2d}
C.~Albertsson, U.~Lindstr\"om and M.~Zabzine,
{\it N = 1 supersymmetric sigma-model with boundaries. I},
Commun.\ Math.\ Phys.\  {\bf 233} (2003) 403
[arXiv:hep-th/0111161];
{\it N = 1 supersymmetric sigma-model with boundaries. II},
Nucl.\ Phys.\ B {\bf 678} (2004) 295
[arXiv:hep-th/0202069].



\bibitem{Koerb}
  P.~Koerber, S.~Nevens and A.~Sevrin,
  {\it Supersymmetric non-linear sigma-models with boundaries revisited,}
  JHEP {\bf 0311} (2003) 066
  [arXiv:hep-th/0309229].
 
 \bibitem{Hori}
  K.~Hori, A.~Iqbal and C.~Vafa,
  {\it D-branes and mirror symmetry,}
  arXiv:hep-th/0005247; 
  K.~Hori,
  {\it Linear models of supersymmetric D-branes,}
  arXiv:hep-th/0012179.
  
  
\bibitem{Warner}
N.~P.~Warner,
{\it Supersymmetry in boundary integrable models},
Nucl.\ Phys.\ B {\bf 450} (1995) 663
[arXiv:hep-th/9506064].

\bibitem{HW}
P.~Ho\v{r}ava and E.~Witten,
{\it Heterotic and type I string dynamics from eleven dimensions,}
  Nucl.\ Phys.\ B {\bf 460} (1996) 506
  [arXiv:hep-th/9510209]; 
{\it Eleven-Dimensional Supergravity on a Manifold with Boundary},
Nucl.\ Phys.\ B {\bf 475}, 94 (1996)
[arXiv:hep-th/9603142].

\bibitem{BDSBM}
A. Bilal, J.-P. Derendinger and R. Sauser,
{\it M-theory on $S^1/Z_2$ : New Facts from a Careful Analysis}, 
Nucl. Phys. {\bf B576} (2000) 347-374 [arXiv:hep-th/9912150];
A.~Bilal and S.~Metzger,
{\it Anomaly cancellation in M-theory: a critical review},
Nucl. Phys. {\bf B675} (2003) 416-446,
[arXiv:hep-th/0307152].

\bibitem{D5}
E.~A.~Mirabelli and M.~E.~Peskin,
{\it Transmission of supersymmetry breaking from a 4-dimensional 
boundary},
Phys.\ Rev.\ D {\bf 58} (1998) 065002
[arXiv:hep-th/9712214];
A.~Lukas, B.~A.~Ovrut, K.~S.~Stelle and D.~Waldram,
 {\it The Universe as a domain wall,}
  Phys.\ Rev.\ D {\bf 59} (1999) 086001
  [arXiv:hep-th/9803235] and 
{\it Heterotic M-theory in five dimensions,}
  Nucl.\ Phys.\ B {\bf 552} (1999) 246
  [arXiv:hep-th/9806051]; 
  A.~Falkowski, Z.~Lalak and S.~Pokorski,
 {\it Supersymmetrizing branes with bulk in five-dimensional supergravity,}
  Phys.\ Lett.\ B {\bf 491} (2000) 172
  [arXiv:hep-th/0004093];
E. Bergshoeff, R. Kallosh and A. Van Proyen, {\it Supersymmetry in 
singular spaces}, JHEP {\bf 0010} (2000) 033, [arXiv:hep-th/0007044];
N.~Arkani-Hamed, T.~Gregoire and J.~G.~Wacker,
 {\it Higher dimensional supersymmetry in 4D superspace,}
  JHEP {\bf 0203} (2002) 055
  [arXiv:hep-th/0101233];
J.~Bagger, F.~Feruglio and F.~Zwirner,
  {\it Brane induced supersymmetry breaking,}
  JHEP {\bf 0202} (2002) 010
  [arXiv:hep-th/0108010];
  A.~Hebecker,
 {\it 5D super Yang-Mills theory in 4-D superspace, 
superfield brane  operators,
 and applications to orbifold GUTs,}
  Nucl.\ Phys.\ B {\bf 632} (2002) 101
  [arXiv:hep-ph/0112230];
   T.~Kugo and K.~Ohashi,
  Prog.\ Theor.\ Phys.\  {\bf 108} (2002) 203
  [arXiv:hep-th/0203276];
M.~Zucker,
   {\it Off-shell supergravity in five-dimensions and supersymmetric brane world
  scenarios,}
  Fortsch.\ Phys.\  {\bf 51} (2003) 899;
D.~V.~Belyaev,
{\it Boundary conditions in supergravity on a manifold with boundary},
JHEP {\bf 0601} (2006) 047
[arXiv:hep-th/0509172].

\bibitem{Erd}
O.~DeWolfe, D.~Z.~Freedman and H.~Ooguri,
  {\it Holography and defect conformal field theories,}
  Phys.\ Rev.\ D {\bf 66} (2002) 025009
  [arXiv:hep-th/0111135]; 
  J.~Erdmenger, Z.~Guralnik and I.~Kirsch,
  {\it Four-dimensional superconformal theories with interacting boundaries or
  defects,}
  Phys.\ Rev.\ D {\bf 66} (2002) 025020
  [arXiv:hep-th/0203020];
  E.~D'Hoker, J.~Estes and M.~Gutperle,
  {\it Interface Yang-Mills, supersymmetry, and Janus,}
  arXiv:hep-th/0603013,
  and references therein.

\bibitem{KR}
A.~Karch and L.~Randall,
{\it Open and closed string interpretation of SUSY CFT's on branes with boundaries,}
JHEP {\bf 0106} (2001) 063
[arXiv:hep-th/0105132];
C.~Bachas and M.~Petropoulos,
{\it Anti-de-Sitter D-branes,}
JHEP {\bf 0102} (2001) 025
[arXiv:hep-th/0012234].

\bibitem{BDDO}
C.~Bachas, J.~de Boer, R.~Dijkgraaf and H.~Ooguri,
{\it Permeable conformal walls and holography},
JHEP {\bf 0206} (2002) 027
[arXiv:hep-th/0111210].

\bibitem{PvN}
P.~van Nieuwenhuizen and D.~V.~Vassilevich,
{\it Consistent boundary conditions for supergravity},
Class.\ Quant.\ Grav.\  {\bf 22} (2005) 5029
[arXiv:hep-th/0507172];
D.~V.~Belyaev and P.~van Nieuwenhuizen,
{\it Simple d=4 supergravity with a boundary},
JHEP {\bf 0809} (2008) 069 
[arXiv:0806.4723 [hep-th]].

\bibitem{Esposito}
G.~Esposito, A.~Y.~Kamenshchik and G.~Pollifrone,
{\it Euclidean quantum gravity on manifolds with boundary,}
Fundam.\ Theor.\ Phys.\  {\bf 85} (1997) 1;
G.~Esposito and A.~Y.~Kamenshchik,
{\it One-loop divergences in simple supergravity: boundary effects},
Phys.\ Rev.\  D {\bf 54} (1996) 3869
[arXiv:hep-th/9604182];
G.~Esposito,
{\it Local boundary conditions in quantum supergravity},
Phys.\ Lett.\  B {\bf 389} (1996) 510
[arXiv:hep-th/9608076];
G.~Esposito and G.~Pollifrone,
{\it Noncovariant gauges in simple supergravity},
Int.\ J.\ Mod.\ Phys.\  D {\bf 6} (1997) 479
[arXiv:hep-th/9701176].

\bibitem{PvN2}
D.~V.~Belyaev and P.~van Nieuwenhuizen,
{\it Rigid supersymmetry with boundaries},
JHEP {\bf 0804}, 008 (2008)
[arXiv:0801.2377 [hep-th]].

\bibitem{GaiottoW}
D.~Gaiotto, E.~Witten,
{\it Supersymmetric Boundary Conditions in N=4 Super Yang-Mills Theory},
[arXiv:0804.2902 [hep-th]].

\bibitem{abgif}
A. Bilal, {\it Introduction to supersymmetry},
Lecture notes ``Gif2000'' [archive:hep-th/0101055];
J.-P. Derendinger, unpublished lecture notes.

\bibitem{WB}
J. Wess and J. Bagger, {\it Supersymmetry and Supergravity},
Princeton University Press, (1983), (second edition: 1992).

\bibitem{NdW}
B.~de Wit, A.~K.~Tollsten and H.~Nicolai,
{\it Locally supersymmetric D = 3 nonlinear sigma-models},
Nucl. Phys. {\bf B392} (1993) 3
[arXiv:hep-th/9208074].


\end{thebibliography}
\end{document}